\documentclass[%
 reprint,
superscriptaddress,
 amsmath,amssymb,
 aps,
]{revtex4-2}

\usepackage{placeins}

\usepackage{graphicx}
\usepackage{dcolumn}
\usepackage{bm}
\usepackage{braket}
\usepackage{color}
\usepackage{amsthm}
\usepackage{dsfont}
\usepackage[absolute]{textpos}
\usepackage{xcolor}
\usepackage{fancyhdr}
\usepackage{listings}
\usepackage{algorithmic}
\usepackage[ruled,vlined]{algorithm2e}
\usepackage{listings}
\usepackage{xcolor}

\definecolor{codegreen}{rgb}{0,0.6,0}
\definecolor{codegray}{rgb}{0.5,0.5,0.5}
\definecolor{codepurple}{rgb}{0.58,0,0.82}
\definecolor{backcolour}{rgb}{0.95,0.95,0.92}

\lstdefinestyle{mystyle}{
    backgroundcolor=\color{backcolour},   
    commentstyle=\color{codegreen},
    keywordstyle=\color{magenta},
    numberstyle=\tiny\color{codegray},
    stringstyle=\color{codepurple},
    basicstyle=\ttfamily\footnotesize,
    breakatwhitespace=false,         
    breaklines=true,                 
    captionpos=b,                    
    keepspaces=true,                 
    numbers=left,                    
    numbersep=5pt,                  
    showspaces=false,                
    showstringspaces=false,
    showtabs=false,                  
    tabsize=2
}

\DeclareMathOperator{\Tr}{Tr}

\def\bx{\mathbf{x}}

\def\balpha{\bm{\alpha}}

\usepackage{appendix}

\preprint{APS/123-QED}

\begin{document}
\title{Data-Centric Machine Learning in Quantum Information Science}

\author{Sanjaya Lohani}
\email[]{slohan3@uic.edu}
\affiliation{Dept.~of Electrical \& Computer Engineering, University of Illinois Chicago, Chicago, IL 60607, USA}
\author{Joseph M. Lukens}
\affiliation{Quantum Information Science Section, Oak Ridge National Laboratory, Oak Ridge, TN 37831, USA}
\author{Ryan~T. Glasser}
\affiliation{Tulane University, New Orleans, LA 70118, USA}
\author{Thomas~A. Searles}
\email[]{tsearles@uic.edu}
\affiliation{Dept.~of Electrical \& Computer Engineering, University of Illinois Chicago, Chicago, IL 60607, USA}
\author{Brian~T. Kirby}
\email[]{brian.t.kirby4.civ@army.mil}
\affiliation{Tulane University, New Orleans, LA 70118, USA}
\affiliation{DEVCOM US Army Research Laboratory, Adelphi, MD 20783, USA}
\date{\today}


\begin{abstract}
We propose a series of data-centric heuristics for improving the performance of machine learning systems when applied to problems in quantum information science. In particular, we consider how systematic engineering of training sets can significantly enhance the accuracy of pre-trained neural networks used for quantum state reconstruction without altering the underlying  architecture. We find that it is not always optimal to engineer training sets to exactly match the expected distribution of a target scenario, and instead, performance can be further improved by biasing the training set to be slightly more mixed than the target. This is due to the heterogeneity in the number of free variables required to describe states of different purity, and as a result, overall accuracy of the network improves when training sets of a fixed size focus on states with the least constrained free variables. For further clarity, we also include a  ``toy model'' demonstration of how spurious correlations can inadvertently enter synthetic data sets used for training, how the performance of systems trained with these correlations can degrade dramatically, and how the inclusion of even relatively few counterexamples can effectively remedy such problems.  
\end{abstract}
\maketitle
\begin{textblock}{13.3}(1.4,15)
\noindent\fontsize{7}{7}\selectfont \textcolor{black!30}{This manuscript has been co-authored by UT-Battelle, LLC, under contract DE-AC05-00OR22725 with the US Department of Energy (DOE). The US government retains and the publisher, by accepting the article for publication, acknowledges that the US government retains a nonexclusive, paid-up, irrevocable, worldwide license to publish or reproduce the published form of this manuscript, or allow others to do so, for US government purposes. DOE will provide public access to these results of federally sponsored research in accordance with the DOE Public Access Plan (http://energy.gov/downloads/doe-public-access-plan).}
\end{textblock}

\section{Introduction}\label{intro}

Machine learning (ML) is quickly becoming a standard tool for approaching and analyzing problems in quantum information science (QIS).
Recent applications include state classification \cite{lu2018separability, harney2020entanglement,ahmed2021classification, wu2021machine}, quantum control \cite{niu2019universal,zhang2019does,porotti2019coherent,bukov2018reinforcement,ding2021breaking}, sensing \cite{ban2021neural,xu2019generalizable,schuff2020improving}, parameter estimation for deployed systems \cite{wang2019machine,ding2020predicting}, turbulence correction \cite{lohani2018use,lohani2018turbulence,knutson2016deep,bhusal2021spatial, lohani2020generative}, and state reconstruction \cite{ahmed2021quantum}, among many others \cite{carleo2019machine,dunjko2018machine,bharti2020machine, lohani2020coherent,PhysRevLett.127.260401}.
Although the motivations for adopting ML in the QIS context vary, they are often related to the ability of ML systems to perform optimization tasks in highly constrained or non-convex situations and the potential improvements in resource scaling compared to standard techniques.  

Efforts to improve the performance of ML systems are generally classified as either model-centric or data-centric.
Model-centric techniques focus on altering the underlying architecture of an ML system.  
Examples include increasing the number of hidden layers in a deep neural network, tailoring the structure of a model, modifying the loss function \cite{genois2021quantum} or tweaking the reward function in reinforcement learning.  
Alternatively, data-centric \cite{huang2021power, liu2021autodc, motamedi2021data, northcutt2021confident, pleiss2020identifying,westermann2021data, hajij2021data, whang2021data, lee2021augment, paiva2021pyhard} methods---leaving the system's architecture unchanged---endeavor to improve system performance by using enhanced data sets, e.g., removing spurious correlations, increasing the accuracy of labels, increasing the variety of sampled situations covered by the data, or distilling the data sets to improve efficiency \cite{wang2018dataset}. 

Given the relative maturity and availability of ML models and systems, and how similarly many state-of-the-art models perform \cite{lucic2017gans}, it has been suggested that data-centric techniques represent an undervalued opportunity to boost system performance \cite{sambasivan2021everyone}.
This recommendation is especially relevant to domain scientists deploying ML in their particular field of research where model-centric methods may be outside of their expertise.
In other words, applying domain-specific knowledge to improve the quality and accuracy of a data set is likely the most efficient and direct route to performance improvements for those not specializing in ML specifically.  

Data-centric ML techniques have found wide applicability in a variety  of domains  including legal \cite{westermann2021data}, natural language processing \cite{xu2021dataclue}, image classification \cite{lee2021augment}, and medical prognosis \cite{paiva2021pyhard}.
In the context of QIS, a data-centric approach to improving state reconstruction accuracy was implemented where expected statistical and experimental noise were included into training sets of a convolutional neural network (CNN), resulting in overall performance improvements \cite{lohani2020machine,danaci2021machine,lohani2021experimental}.
Beyond the inclusion of artificial errors, it was also recently shown that even very general pieces of prior information in the construction of training sets---such as the expected mean purity---can improve the performance of ML-based state reconstruction systems \cite{lohani2021improving}.

\begin{figure*}[ht]
\centering
\includegraphics[width=.9\linewidth]{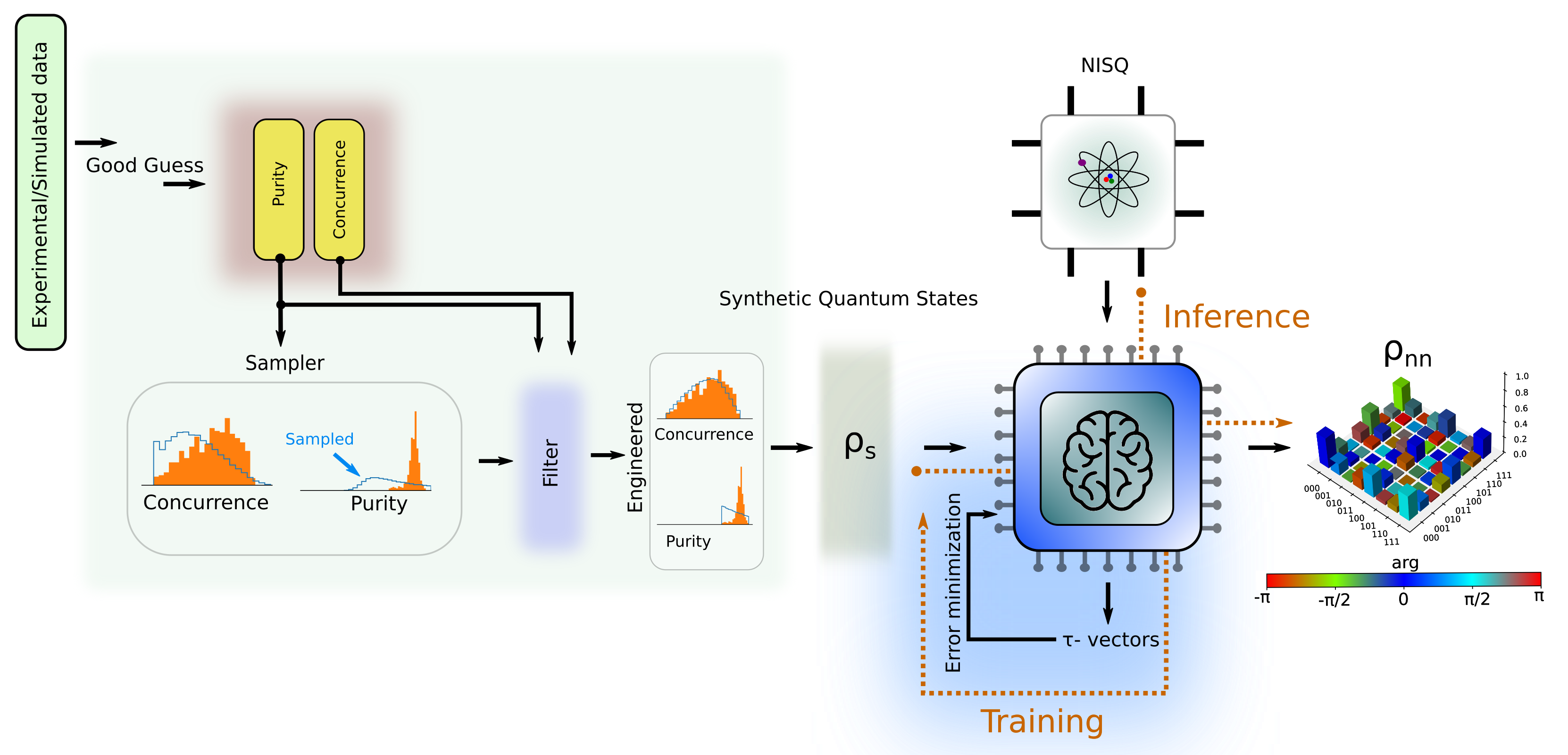}
\caption{\textbf{A schematic of a data-centric ML approach in QIS.} Inputs are coarse properties of the distributions expected in our test system, e.g., the limits (maximum and minimum) of the purity and concurrence. To obtain the training set, we sample from a suitably chosen distribution of random quantum states, and then apply a bandpass filter to eliminate any samples lying outside of the specified extrema. 
These engineered training sets are used in optimizing the trainable parameters of an ML model, as indicated by a dotted brown arrow labelled ``Training.'' Finally, the resulting neural network is enlisted to infer unknown quantum states probed by measurements on NISQ-era hardware (``Inference'' arrow). An example reconstructed three-qubit quantum state is shown at the right, where shading denotes the phase.}
\label{fig:intro}
\end{figure*}

This paper develops data-centric heuristics specifically targeting classical ML applications in QIS.
For concreteness, we demonstrate the effectiveness of the heuristics using a pre-trained CNN-based quantum state reconstruction system.
However, the heuristics themselves are based on the general properties of distributions of quantum states and are use-case agnostic.  
After reviewing the architecture of our CNN in Sec.~\ref{NN}, we begin in Sec.~\ref{Spurious} by introducing a representative example where spurious correlations between purity and entanglement in a training set cause our network to misclassify separable pure states as entangled.
While this ``toy model" intuitively demonstrates the impact of training set construction on application performance, it also leads to our first heuristic: engineering data sets to include even comparatively few counterexamples is sufficient to remedy errors due to spurious correlations.

In Sec.~\ref{V}, we present three more data-centric approaches that improve the reconstruction fidelity of our system.  
We begin by reviewing various distributions of random quantum states used to generate synthetic training sets for ML systems and describe a method for further engineering them to incorporate prior information using simple ``bandpass filters."
To illustrate the utility of our training set engineering approach, we perform state reconstruction on randomly sampled states from a cloud-accessed 7-qubit quantum processor and demonstrate fidelity improvements.
To stress the data-centric nature of these fidelity improvements, we compare our results to those achievable using the same datasets but with a model-centric approach that increases the CNN's trainable parameters.
Then, in Sec.~\ref{shotNoise}, we build on prior results found in \cite{lohani2020machine,danaci2021machine,lohani2021experimental} and demonstrate that synthetically incorporating statistical noise comparable to that found in a test set into the training set can result in significant performance improvements. 
Finally, in Sec.~\ref{heterogeneity}, we consider situations where states covering a wide range of purity values are expected in the target distribution of an ML system.
We find that, in this case, it is not always optimal to train on the exact distribution expected, and instead, it is more efficient to bias training sets towards the more mixed end of the distribution.
We argue that this surprising result is due to the heterogeneity of the free variables in states of differing purity.

\section{Neural network} \label{NN}
We implement a custom-designed CNN that takes tomographic measurement values as inputs and reconstructs an estimate of the density matrix as the output, similar to systems described in \cite{lohani2021improving,lohani2021experimental,lohani2020machine}.   
Our system has  a convolutional layer with a kernel size of (2, 2), stride lengths of 1, ReLU as an activation function, and filters of size 25. Then we add a max-pooling layer with a pool-size of (2, 2), stride lengths of 2, and a ``valid'' padding, followed by a flattening layer. 
Next, we attach a fully connected dense layer ($dense\_1$) using the ReLU activation function. Then, we apply a dropout layer with a 50$\%$ dropout rate, followed by another fully connected dense layer ($dense\_2$), again, using the ReLU activation function, followed by a dropout layer with the same rate. After this, we attach another fully connected dense layer ($dense\_3$) with a linear activation. Note that the number of trainable parameters  depends upon the number of neurons at the $dense\_1$, $dense\_2$, $dense\_3$ layers, and the number of qubits. 

The output of $dense\_3$ is a vector $\tau_{nn}$ that  
defines a corresponding density matrix through the Cholesky decomposition. In general, any density matrix can be written as $\rho\,=\,\zeta(\tau)\zeta(\tau)^\dagger$, where $\zeta(\tau)$ is a lower triangular matrix. The nonzero elements of the matrix $\zeta(\tau)$ can be rearranged into a vector as given by
\begin{equation}
    \zeta(\tau) \longrightarrow (\tau_0,\, \tau_1,\, \tau_2,\, ...., \, \tau_{2^{2d}-1}),
\end{equation}
where $d$ is the number of qubits. The first $2^d$ elements represent the diagonal entries, and the remaining components populate the real and imaginary parts of the off-diagonal entries. As an example, in the two-qubit case $\zeta(\tau)$ is given by
\begin{equation}
\begin{aligned}
    &\zeta(\tau)\,=\,
        \begin{bmatrix}
        \tau_0 & 0& 0& 0\\
        \tau_4+i\tau_5 & \tau_1 &0 &0\\
        \tau_{10}+i\tau_{11} & \tau_6+i\tau_7 &\tau_2 &0\\
        \tau_{14}+i\tau_{15} & \tau_{12}+i\tau_{13} &\tau_8+i\tau_9 &\tau_3\\
        \end{bmatrix}. \\
\end{aligned}
\label{eqn:cholesky_T}
\end{equation}

During training, the ground truth target vector $\tau_g$ is provided to the network, and the trainable parameters are optimized to minimize the mean squared error (MSE) $\braket{||\tau_{nn}-\tau_{g}||^2}$, where the average is taken over the full training set. Once trained, the network takes any collection of measurement values as an input and outputs a prediction, $\tau_{nn}$.
For validation of the trained network, we utilize measurement data generated from a test set of density matrices $\rho_g$---which may not match the training set---and compute the density matrix $\rho_{nn}$ corresponding to the network output $\tau_{nn}$. The fidelity $F(\rho_{nn},\rho_g) = \Big[ \Tr\sqrt{\sqrt{\rho_{g}}\rho_{nn}\sqrt{\rho_{g}}}\Big]^2$ is then used to quantify accuracy.

A schematic summarizing our approach for engineering distributions is shown in Fig.~\ref{fig:intro}. 
The inputs to our system are the ranges of the desired purity and concurrence distributions, but generally, any quantifiable property of a distribution can be substituted.  
In our particular case, we estimate the approximate range of the distribution of states generated from the IBMQ as in \cite{lohani2021experimental}, but many other techniques can be used without requiring full state reconstruction \cite{huang2020predicting,lukens2021bayesian}. 
The input purity and concurrence ranges inform the selection of an initial distribution of random quantum states, which is passed through a  simultaneous bandpass filter in both purity and concurrence, resulting in our final engineered training set. A detailed description of the engineering procedure is presented in Sec.~\ref{V}.

We then use these engineered training sets to optimize the trainable parameters of an ML model as previously mentioned, which is shown by a dotted brown arrow ``Training'' in Fig. \ref{fig:intro}. Once trained, the resulting neural network is, finally, enlisted to infer unknown quantum states probed by measurements on noisy intermediate-scale quantum (NISQ) hardware, in our particular case the IBMQ, as indicated by an ``Inference'' arrow.


\section{Spurious correlations and lack of variation}
\label{Spurious}
As an illustrative example to introduce our data-centric approach to quantum state tomography (QST), we first examine an ML system trained on a set of density matrices containing perfect correlations between purity and entanglement. 
We then study the performance of this system when used as a separability-entanglement classifier on generic states that are counterexamples to the learned correlation.
In this sense, how spurious correlations impact reconstruction fidelity and separability-entanglement classification is related to the ML concept of generalizability, which considers how well a system will perform on data not included in its training set~\cite{kawaguchi2017generalization,miller2021accuracy}.  
We will show that our network indeed learns the correlations between purity and entanglement present in the training set, which limits generalizability and  results in a high error rate when classifying pure states as separable or entangled.  
Yet we will then demonstrate that including only a modest number of counterexamples in the training set can significantly mitigate this issue---a paradigmatic ``data-centric'' improvement.

\begin{figure}[tb!]
\centering
\includegraphics[width=\linewidth]{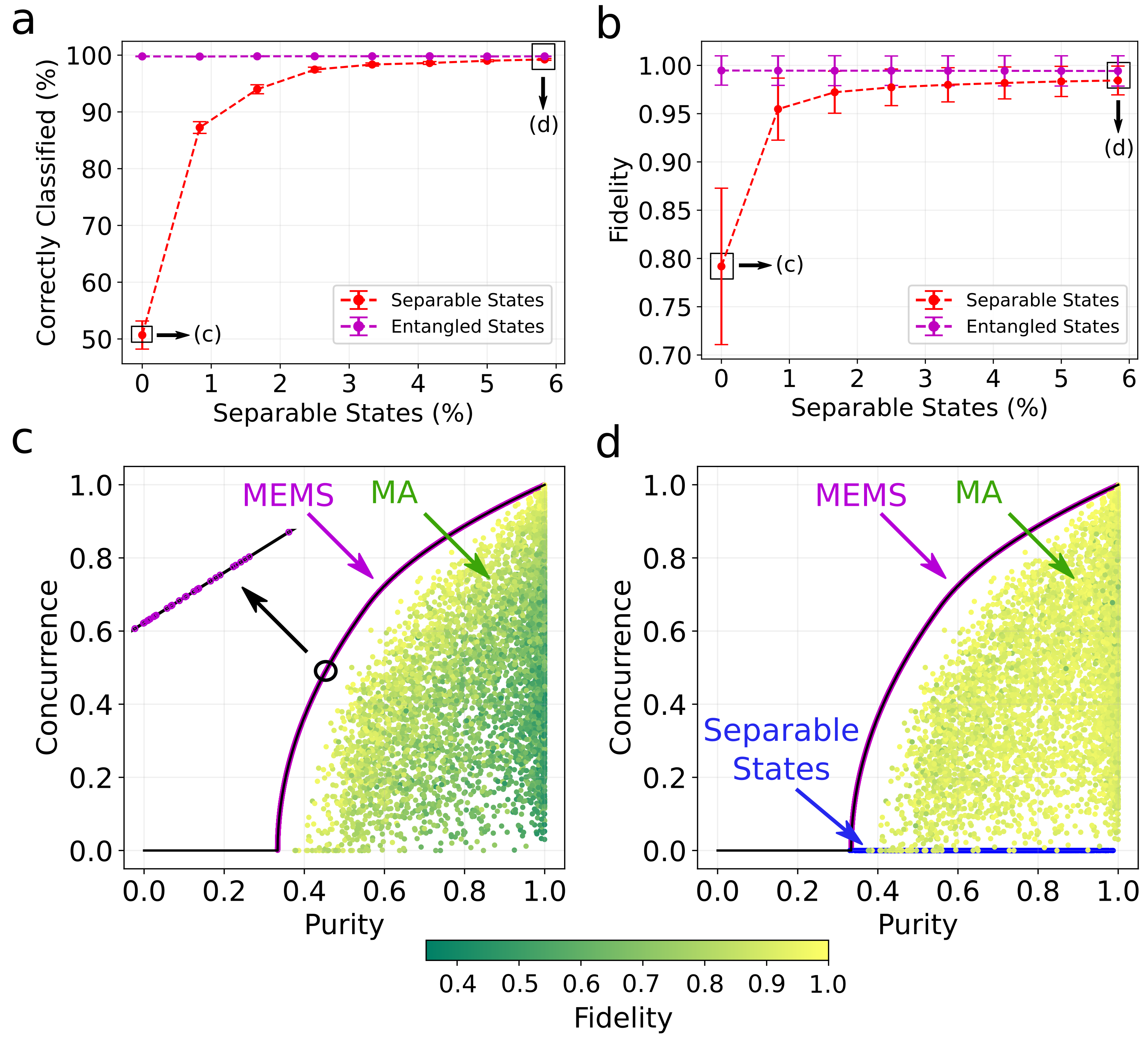}
\caption{\textbf{Reducing spurious correlations.} (a) Accuracy of entanglement-separability classification and (b) network reconstruction fidelity versus the percentage of separable states added to a training set containing entangled states. Reconstruction fidelity for test states from the MA distribution for (c) a MEMS-only trained network and (d) after adding a small fraction of separable states into the training set. MEMS (separable) states are shown by magenta (blue) dots. A small portion of the MEMS line is magnified (as indicated by a small circle) and shown in the top-left inset. The generated test sets with MA distributions are shown by the inner dots, while their corresponding reconstruction fidelities are indicated by a color bar. The rectangular boxes in (a,b) represent the pre-trained models that are used for results shown in the corresponding plot (c,d) below.}
\label{fig:spurious_correlationi}
\end{figure}

To define the restricted subspace within the overall Hilbert space that enforces a strong correlation between purity and entanglement, we generate our training states from local rotations of two-qubit maximally entangled mixed states (MEMS).
The MEMS define a particular class of states which, for a given linear entropy, have the maximum possible concurrence \cite{munro2001maximizing,wei2003maximal}. 
In general, MEMS can be expressed as (up to local rotations) 
\begin{equation}
    \begin{aligned}
    &\rho_{MEMS}\,=\,
        \begin{bmatrix}
        g(\gamma)& 0& 0&\frac{\gamma}{2}\\
        0 & 1 - 2g(\gamma)& 0 &0 \\
        0& 0 &0 &0\\
        \frac{\gamma}{2}& 0 &0 & g(\gamma)
        \end{bmatrix}, \\
\end{aligned} 
\label{eq:MEMS_DEF}
\end{equation}
where
\begin{equation}
     g(\gamma) = \begin{cases}
     \frac{\gamma}{2} & ; \;\; \gamma \geq \frac{2}{3} \\    
     \frac{1}{3} & ; \;\; \gamma < \frac{2}{3}\\
    \end{cases},
\label{eq: MEMS_rho}
\end{equation}
and the parameter $\gamma\in[0,1]$ is equal to the concurrence. For consistency of presentation with results later in this manuscript, we consider the purity of the MEMS instead of the linear entropy.  The purity of the state in Eq.~\eqref{eq:MEMS_DEF} is given by
\begin{equation}
    P(\gamma)=1-4g(\gamma)+6g^2(\gamma)+\frac{\gamma^{2}}{2},
    \label{eq:MEMS_PURITY}
\end{equation}
which ranges from $\frac{1}{3}\le P(\gamma)\le 1$.
We generate an element of our training set $\rho_{MEMS}'$ according to
\begin{equation}
    \rho'_{MEMS} = \left(U_{a}(2)\otimes U_{b}(2)\right)\rho_{MEMS} \left(U_{a}(2)\otimes U_{b}(2)\right)^{\dagger},
\label{eq: MEMS_rotations}
\end{equation}
where $\gamma$ is drawn from the Uniform distribution, $\gamma \thicksim \mathrm{Uniform}(0, \, 1)$, and $U_{i}(2)$ is a two-dimensional Haar-random unitary matrix applied to qubit $i$.
When plotting the concurrence of these states as a function of their purity, they form a curve as shown in Fig.~\ref{fig:spurious_correlationi}(c,d).
Note that the MEMS span a wide range of possible values of purity and concurrence.
We stress this fact about the MEMS to demonstrate how even for two qubits, the simplest of all entangled systems, it can be challenging to detect spurious correlations when only considering general properties of states independent of each other.
In this case, the relationship between purity and entanglement \cite{munro2001maximizing,wei2003maximal} is well known, but such relationships may be significantly more difficult to detect in more complex systems.

Due to the correlation between purity and entanglement in the training set, our network has never been exposed to separable states of high purity.  
Therefore, to understand the effects of these correlations we will use our network to reconstruct randomly generated separable states with $P > \frac{1}{3}$ and classify them as separable or entangled based on the Peres--Horodecki positive partial transpose (PPT) criterion \cite{PhysRevLett.77.1413, Horodecki1996}.
We generate these states according to
\begin{equation}
    \rho_{s}\,=\,\rho_a \otimes \rho_b, \quad \textrm{such that}\quad \Tr(\rho^2_{s}) > \frac{1}{3},
    \label{eq: rho_s}
\end{equation}
where $\rho_a$ and $\rho_b$ are random full-rank density matrices sampled from the Hilbert--Schmidt (HS) distribution.  The use of the HS distribution here does not meaningfully impact the results, and details related to the distribution are described in Sec.~\ref{distributions}.
We find that our network has significant error in this scenario and fails to correctly classify approximately $50$\% of the states from $\rho_{s}$, as shown by the first data point in Fig.~\ref{fig:spurious_correlationi}(a), suggesting that the impact of spurious correlations on the performance of ML-based QST can be dramatic. 

We next investigate how these errors can be mitigated through the inclusion of a modest number of counterexamples in the training set.  In particular, we include randomly sampled states from $\rho_{s}$ in the initial training set, such that the total number of states in the training set (states from $\rho_{MEMS}'$ and $\rho_{s}$) is 
\begin{equation}
    N_\mathrm{train} = N' + N_s,
\end{equation}
where $N_\mathrm{train}$ represents the total number of training states, $N'$ is the number of states drawn using Eq.~\eqref{eq: MEMS_rotations} and $N_s$ the total number sampled from Eq.~\eqref{eq: rho_s}. We fix $N_\mathrm{train}=30,000$ and modify the fraction $N_s/N_\mathrm{train}$. 
For the test set, we sample 5000 random states using Eq.~\eqref{eq: rho_s}. Note that the training and test sets are drawn randomly and independently.

We vary $N_s$ from 0 to 1750 with a step size of 250 and train a separate neural network at $dense\_1\,=\,3050$ and $dense\_2\,=\,1650$ up to 400 epochs at a learning rate of 0.008 for each of them as previously described. 
After this, the pre-trained networks are again used to classify states from $\rho_{s}$ as separable or entangled.  
The state classification accuracy versus the percentage of separable states added to a training set is shown in Fig.~\ref{fig:spurious_correlationi}(a). 
Note that we train the same network architecture 10 times for each case and take the average of all the predictions (from 10 trials) for the given state to minimize the effects of random initialization during training. 
The $N_s=0$ point on the curve represents the result described above where the entire training set is drawn from states in $\rho_{MEMS}'$.
We see that with an increasing percentage of separable states in the training set, the network state classification accuracy increases rapidly. 

Similarly, we also measure the reconstruction fidelities for states from $\rho_{s}$, with the results shown as the dotted red line in Fig.~\ref{fig:spurious_correlationi}(b). 
As expected, with the additional separable state examples in the training set, we have significantly enhanced the network's reconstruction fidelity for separable states. 
Furthermore, in order to cross-check if the network performance for entangled states has been affected by the modified training sets, we again independently sample 5000 random states according to Eq.~\eqref{eq: MEMS_rotations}.
Note that the purity of each $\rho'_{MEMS}$ generated according to Eq.~\eqref{eq: MEMS_rotations} will always be greater than $\frac{1}{3}$, and that except for the vanishingly small case where $\gamma=0$, the sampled states are entangled. 
The reconstruction fidelity for the sampled $\rho'_{MEMS}$ test states are shown as the magenta lines in Fig.~\ref{fig:spurious_correlationi}(a,b).
The error bars show one standard deviation from the mean.
The fidelity for this test set remains high and constant for all examined training sets.
Hence, we find that the inclusion of a small number of counterexamples in the training set significantly improves classification performance on out-of-correlation states without reducing overall network performance.

To further illustrate the overall impact of spurious correlation and the effect of the mitigation strategy, we also reconstruct states from across the purity-concurrence plane with a network trained only on the MEMS class ($N_s/N_\mathrm{train}=0$) and a network trained on counterexamples as well ($N_s/N_\mathrm{train}=0.058$), the results of which are plotted in Figs.~\ref{fig:spurious_correlationi}(c,d).
Each point represents a randomly sampled density matrix colored by the reconstruction fidelity.
Figure~\ref{fig:spurious_correlationi}(c) corresponds to the same network as indicated by the leftmost boxes in Fig.~\ref{fig:spurious_correlationi}(a,b), and Fig.~\ref{fig:spurious_correlationi}(d) to the rightmost boxes.  
The figure includes 5000 randomly sampled points from the Mai--Alquier (MA) distribution [see Sec.~\ref{distributions}] with $\alpha=0.1$, chosen for illustrative purposes.  The reconstructed fidelities for the test states are represented by a color bar. The magenta and blue dots, respectively, represent the MEMS and added separable states, where the inner dots indicate the MA-distributed states.

As expected, in Fig.~\ref{fig:spurious_correlationi}(c) separable states of high purity (residing toward the lower right hand corner) are reconstructed with low fidelity, since these states possess the most extreme deviation from the correlation found in MEMS. In contrast, we find a significant enhancement for the network trained with MEMS and a few examples of separable states, as shown in Fig.~\ref{fig:spurious_correlationi}(d). 
Interestingly, although the improvement is most pronounced for separable pure states, the modified training set increases reconstruction fidelity across the entire purity-concurrence plane. 

To conclude this section, we note that with respect to the use of MEMS training sets, QIS researchers would be unlikely to expect generalizability, and it is perhaps unsurprising that neural networks trained on them would perform so poorly on other states. Nevertheless, their strong correlations acutely highlight the broad issue of spurious correlations---which in many situations may prove much more difficult to detect---as well as indicate a simple mitigation strategy based on tailored training data. In the following sections, we apply these general ideas to situations of more practical interest in QIS, exploring a variety of density matrix distributions that all offer full Hilbert space support, and compare their performance in ML-based QST. In these more nuanced cases, we again will find noticeable improvements with engineered training sets, for multiple experimentally relevant contexts.

\section{Engineering training sets to improve reconstruction fidelity}\label{V}

\subsection{Distributions of random quantum states}
\label{distributions}

In this subsection, we briefly review the salient features of the most common methods for defining distributions of random quantum states.
Beyond fundamental motivations \cite{zyczkowski1999volume,zyczkowski2005average,zyczkowski2001induced}, many efforts to perform state reconstruction and classification using ML-based methods have relied on these distributions to generate training sets \cite{lohani2020machine,lohani2021experimental,lohani2021improving,danaci2021machine,lu2018separability}. 
These distributions will serve as baselines for evaluating ML-based system performance for which our data-centric heuristics will be compared.  
In other words, we aim to improve the performance of ML-based techniques in QIS beyond what is possible from training on these standard distributions.  

\textbf{Hilbert--Schmidt (HS) distribution:}
Random quantum states distributed according to the HS measure can be induced through the partial trace on Haar-random pure states in higher dimensions \cite{zyczkowski2001induced}.
Operationally, ensembles of HS-distributed random quantum states are typically generated by sampling the complex Ginibre ensemble \cite{ginibre1965statistical}, which comprises $D\times D$ complex matrices whose elements are independently drawn from the complex standard normal distribution \cite{zyczkowski2001induced}. 
Specifically, random quantum states distributed according to the HS measure can be obtained using
\begin{equation}
        \rho=\frac{GG^{\dagger}}{\text{Tr}\left(GG^{\dagger} \right)},
    \label{eq:HS}
\end{equation}
where $G$ is a random matrix from the Ginibre ensemble.

\textbf{Bures distribution:} Similar to the case of the HS distribution, a random quantum state $\rho$ from the Bures ensemble can be sampled according to
\begin{equation}
        \rho=\frac{\left(\mathds{1}+U\right)GG^{\dagger}\left(\mathds{1}+U^{\dagger}\right)}{\text{Tr}\left[\left(\mathds{1}+U\right)GG^{\dagger}\left(\mathds{1}+U^{\dagger}\right)\right]},
    \label{eq:Bures}
\end{equation}
where $G$ is, again, a random matrix from the Ginibre ensemble and $U$ is a Haar-distributed random unitary from $U(D)$~\cite{al2010random}.

\textbf{Hilbert--Schmidt--Haar (HS--Haar) distribution:} 
Previous studies have noted that ensembles of random quantum states distributed according to the HS and Bures measures have limited applicability for many NISQ devices due to their low average purities~\cite{lohani2021improving}.
Therefore we define here a simple technique for biasing an arbitrary input distribution toward a higher average purity.
In particular, we consider a convex combination of HS-distributed quantum states ($\rho_{HS}$) and random Haar-distributed pure states ($\rho_H$) as given by
\begin{equation}
    \rho = (1 - \delta) \, \rho_{HS} + \delta \, \rho_H,
\end{equation}
where $\delta$ is chosen uniformly at random from the interval $[0,1]$. 
We will find later in this section that this approach performs surprisingly well in our tests despite its simplicity.

\textbf{Mai--Alquier (MA) distribution:} This distribution was originally studied as a prior for Bayesian QST~\cite{mai2017pseudo,lukens2020practical,lu2020fully,lingaraju2021adaptive,alshowkan2021reconfigurable} and was recently utilized in \cite{lohani2021improving} to generate training sets for ML-based state reconstruction methods.  The MA distribution is defined as a mixture of Haar-random pure  states with coefficients drawn from the Dirichlet distribution. The probability density function of the Dirichlet distribution for vectors $\bx=(x_{1},...,x_{K})$---where the elements of $\bx$ belong to the open $K-1$ simplex  ($x_i\geq 0$ and  $\sum_{i=1}^K x_i = 1$)---is
\begin{equation}
    \text{Dir}(\bx\vert\balpha)=\frac{\Gamma\left(\sum_{i=1}^{K}\alpha_{i}\right)}{\prod_{i=1}^{K}\Gamma(\alpha_{i})}\prod_{i=1}^{K}x_{i}^{\alpha_{i}-1},
    \label{eq:dirichlet_def}
\end{equation}
where $\balpha=(\alpha_{1},...,\alpha_{K})$ with all $\alpha_{i}\ge 0$ defines the concentration parameters and $\Gamma(\cdot)$ is the standard gamma function. The concentration parameters provide flexibility to alter the overall features of the distribution. Therefore, an ensemble of $D$-dimensional mixed states from a convex sum of $K$ Haar-random pure quantum states $\ket{\psi_{i}}$ is written as
\begin{equation}
    \rho=\sum_{i=1}^{K}x_{i}\vert\psi_{i}\rangle\langle\psi_{i}\vert,
    \label{eq:D_state_def}
\end{equation}
where the vector $\bx$ is a random variable distributed according to $\text{Dir}(\bx|\alpha)$: for simplicity, we specialize to the symmetric case $\balpha=\{\alpha,...,\alpha\}$ only. The expectation value of the purity is
\begin{equation}
        \text{E}_{MA}\left[\text{Tr}(\rho^{2})\right]=\frac{D+\alpha(D+K-1)}{D(1+\alpha K)}.
    \label{eq:purity_av1}
\end{equation}
Finally, we note that in \cite{lohani2021improving} strong evidence was presented that the MA distribution reduces to the HS distribution for a specific set of input parameters.

\textbf{\.{Z}yczkowski (Z) distribution:} Dirichlet-distributed vectors $\bx$ are again employed to generate random density matrices as described by \cite{zyczkowski1999volume}, which we refer to as the Z distribution for convenience.
This approach relies on the unitary invariance of the eigenvalues of a density matrix and utilizes the Dirichlet vectors $\bx$ of length $D$ as the eigenvalues of $D$-dimensional states.
Once the eigenvalues are generated they are placed along the diagonal of a $D\times D$ matrix and a Haar-random unitary from $U(D)$ is applied.
Note that the resulting construction is of the same form as Eq.~\eqref{eq:D_state_def} for $K=D$, but with all states in the convex sum orthogonal.
The expectation value of the purity of Z-distributed states is given by
\begin{equation}
    \text{E}_{Z}\left[\text{Tr}\left(\rho^{2}\right)
    \right]=\sum_{j=1}^{D}\text{E}\left[x_{j}^{2}\right]=\frac{1+\beta}{1+D\beta}
\end{equation}
where we have used $\beta$ as the concentration parameter of the Dirichlet distribution so as not to be confused with the MA expressions (where $\alpha$ was used).
Like the MA distribution, the Z distribution can be biased in various ways through manipulation of concentration parameters.  A discussion of how the MA and Z distributions compare against experimentally measured distributions can be found in \cite{lohani2021improving}.
Finally, we note that the Z distribution is a widely employed method for generating random density matrices including for machine-learning applications such as the state classifier described in \cite{lu2018separability}.
\subsection{Generating random test sets on NISQ hardware}
\label{V_A}
In order to illustrate several of our heuristics on data sets representative of real-world experimental scenarios, we will perform quantum state reconstruction and state classification using data obtained from NISQ hardware.  In particular, we utilize data sets consisting of tomographic measurements performed on random quantum states obtained with $ibmq\_jakarta$, one of the IBM Quantum Falcon processors. We first numerically generate 500 Haar-random two-qubit pure states and initialize these on  $ibmq\_jakarta$. Then, the states are automatically transpiled from the backend into the required quantum circuits for generation. The depths of the transpiled quantum circuits---i.e., the longest path from input to output---range between 12 and 16 gates.  
For each state, we perform full state tomography with a total of 36 measurement projections, corresponding to the four outcomes for all 9 two-qubit combinations of the Pauli operators $\{X,Y,Z\}_{1}\otimes\{X,Y,Z\}_{2}$. 

We then reconstruct the measured quantum states using maximum likelihood estimation (MLE) according to the algorithm in~\cite{smolin2012efficient}. Unfortunately, due to random noise on the backend hardware, the estimated states are mixed despite having been programmed as pure \cite{lohani2021experimental}, leaving uncertainty about the ground truth state that the data represent. 
Therefore, to retain the general properties of the distribution generated by $ibmq\_jakarta$  while permitting the construction of test sets with known ground truth states, 
we perform additional rounds of tomographic simulations on the MLE-obtained results; these synthetic measurement results comprise the test sets below.  We simulate measurement results using the methods described in \cite{lohani2021experimental} which further allow us to select the amount of statistical noise (shots) on demand. 

\subsection{Engineering training sets} \label{V_B}

Motivated by the restricted nature of existing techniques, we describe a general method for engineering an arbitrary input distribution to conform to certain general characteristics desired in a training set. 
Several of the previously introduced distributions allow for biasing based on a single input.  This is enough to control, for example, the mean purity~\cite{lohani2021improving}. However, a single parameter may not always be enough to meaningfully constrain a distribution for a given use case. For example, even in the two-qubit case considered here, 
the purity and the entanglement only bound rather than determine each other \cite{munro2001maximizing}. The situation becomes even more complex for higher-dimensional systems where several inequivalent classes of entanglement exist \cite{guhne2009entanglement}.


\begin{table*}[ht]
  \label{tbl:table_plot}
  \includegraphics[width=.85\linewidth]{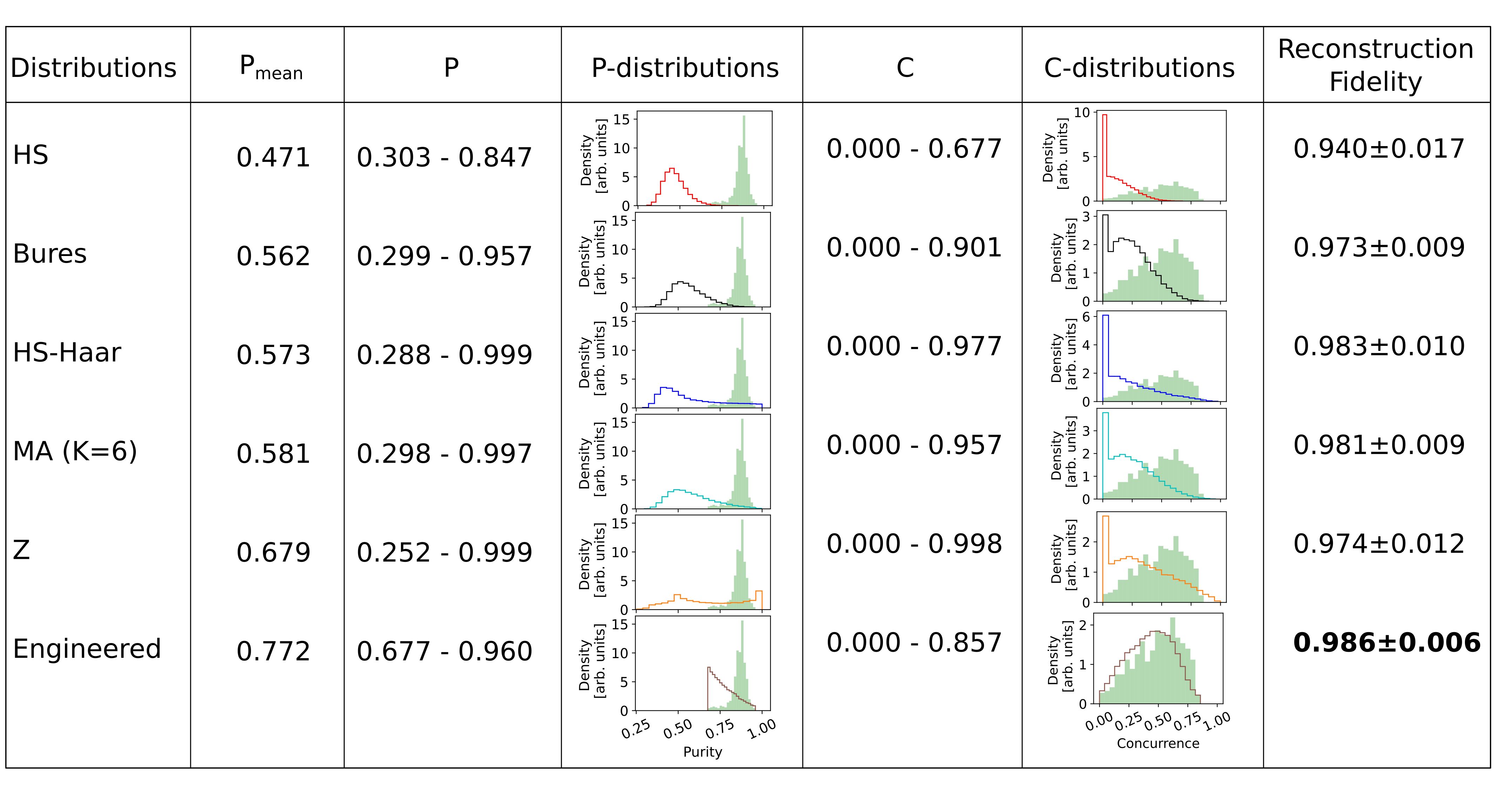}
  \caption{Purity and concurrence distributions of explored training sets. P$_\mathrm{mean}$, P, P-distributions, C, and C-distributions represent the mean purity, range of purity, purity distributions, range of concurrence, and concurrence distributions,  respectively, for training sets generated from the density matrix distributions in the left column. Of particular interest are the ``P-distributions'' and ``C-distributions'' columns which show the NISQ sample distribution in solid green with the training set distributions overlaid. The average reconstruction fidelities for test states from the NISQ distribution are shown in the right column. The error ($\pm$) indicates one standard deviation.}
\end{table*}


The method described in Fig.~\ref{fig:intro} consists of repeatedly sampling from a suitably chosen input distribution followed by the application of a simultaneous bandpass filter for both purity and concurrence. In general, the bandpass filter approach can be applied to any measurable property or properties of the sampled states. However, we have chosen purity and concurrence for this demonstration as they are both well-understood properties of two-qubit density matrices. Hence, in many experiments of interest, the approximate maximum and minimum values of the purity and concurrence are easily inferred.
In short, the bandpass filter approach can be summarized as first randomly sampling states $\rho_{\pi}$ from an arbitrary input distribution $\Pi$ and then passing them through the simultaneous filter given by
\begin{equation}
\begin{aligned}
     \rho_\mathrm{eng} &= \rho_{\pi}\Big[\big(C_{\min}\leq C(\rho_{\pi}) \leq C_{\max}\big)\, \\
    & \&  \big(P_{\min}\leq P(\rho_{\pi}) \leq P_{\max}\big)\Big],
\end{aligned}
\end{equation}
where $\rho_\mathrm{eng}$ are the filtered (engineered) states, and $C$ and $P$ represent the concurrence and purity, respectively. 

In principle, we can use any input distribution $\Pi$ in the above approach. Hence, our method for engineering distributions of random states only requires upper and lower bounds on the purity and concurrence, or some other property of the states, as input. However, the MA distribution is particularly convenient as an initial distribution because it allows extra freedom in biasing the resultant distribution of states. Furthermore, we will find in Sec.~\ref{heterogeneity} that when a test set covers a wide range of purity values, as is often the case in an experimental situation, the optimal training set to maximize mean reconstruction fidelity is more mixed than the test set. 
Therefore, we recommend that input distribution $\Pi$ be chosen according to the following recipe.  First, set $K=D$ (the minimum $K$ value capable of producing full rank matrices) and tune $\alpha$ such that the mean of this distribution is equal to the chosen lower bound.
Second, based on results in Appendix~\ref{tuning_MA}, we then recommend that once $\alpha$ is fixed, the distribution be sampled with $K>D$. We will explore the benefit of this final recommendation more in Sec.~\ref{heterogeneity} and Appendix~\ref{tuning_MA}. Finally, we summarize the full algorithm for engineering distributions in Appendix~\ref{appendix_E}.


\subsection{Data-centric state reconstruction with NISQ hardware} \label{V_C}
We now compare the performance improvements obtainable with both data-centric and model-centric techniques using our ML-based QST system.
Our data-centric methods consist of training our ML-QST system using sets drawn from the various distributions described in Sec.~\ref{distributions} and those engineered using the method in Sec.~\ref{V_B}.
Our model-centric methods consist of increasing the trainable parameters in the network.
Ultimately we will find that data- and model-centric approaches work in a complementary fashion and that our greatest data-centric improvement is found using our engineered distribution approach from Sec.~\ref{V_B}.

The test set used to compare the various trained networks in this section is based on the NISQ-sampled density matrices according to Sec.~\ref{V_A}.
The tomographic data are generated assuming 1024 shots, meaning every Pauli measurement circuit was executed 1024 times each.
The resultant distribution has a purity and concurrence in the range $[0.68,0.96]$ and $[0,0.86]$, respectively, which is used to inform the engineered distribution of Sec.~\ref{V_B}. All 500 generated states are used as a test set. Note that the test states are completely unknown and hidden from the network during training, and the only information used to inform the construction of training sets is the maximum and minimum of the purity and concurrence.

A summary of the distributions used to generate training sets for our ML-QST system is included in Table I.  Of particular interest are the ``P-distributions" and ``C-distributions" columns which show the NISQ sample distributions in solid green with the training set distributions overlaid.  
The $x$-axis in each of these plots is held constant, but the scaling along the $y$-axis is arbitrary.
The HS, Bures, and HS-Haar distributions have no input parameters and hence the $P_\mathrm{mean}$, $P$, and $C$ values list in Table I for these distributions are application-agnostic.
Alternatively, the MA, Z, and engineered distributions all include degrees of freedom that allow for the incorporation of prior information.  
As discussed in Sec.~\ref{V_B}, for the engineered distribution we first select the $\alpha$ value which makes the mean purity of the $K=D$ distribution equal to $P_\mathrm{min}$, and then use a larger $K>D$ to actually generate samples.
In this case we find when $K=4$, the mean of the MA distribution matches $P_\mathrm{min}$ for $\alpha=0.3394$, which we then sample with $K=6$.
These selections are made heuristically with knowledge of $P_\mathrm{min}$ only, but arguments for their selection and considerations of their optimality are considered in Sec.~\ref{heterogeneity} and Appendices~\ref{appendix_C}, \ref{tuning_MA}, and \ref{Only_MA_distributed_results}.
Finally, for a fair comparison of the engineered distribution against the MA and Z distributions we also show the MA distribution with $\alpha=0.3394$ and $K=6$ but without having passed through the bandpass filter and the Z distribution tuned such that its mean purity is equal to $P_\mathrm{min}$.
Although these parameter selections are well motivated by the results in Sec.~\ref{heterogeneity}, we also demonstrate explicitly the impact of this selection in Appendix \ref{appendix_E}, where we see that this choice is near optimal for both the MA and engineered distributions.


For each training set, we  also consider the impact of an additional model-centric approach which consists of increasing the number of trainable parameters in the model. 
As the number of qubits is fixed, the number of neurons in the $dense\_3$ layer is also fixed at 16. Therefore, the number of neurons in the $dense\_1$ and $dense\_2$ layers determine the total number of trainable parameters as shown in Table \ref{table:one}. We use the same network architecture for the model with all combinations of sampling distributions. Additionally, all the networks are trained up to 400 epochs with a learning rate of 0.008. We manually optimize the learning rate, as discussed briefly in Appendix \ref{appendix_D}. 


\begin{table*}[ht]
\centering
 \begin{tabular}{|p{2cm}|p{.75cm}|p{.75cm}|p{.75cm}|p{.75cm}|p{.75cm}|p{.75cm}|p{.75cm}|p{.75cm}|p{.75cm}|p{.75cm}|p{.75cm}|p{.75cm}|p{.75cm}|p{.75cm}|p{.75cm}|} 

 \hline 
 dense\_1 &50&150&250&350&450&550&650&750&850&950&1050&1550&2050&2550&3050 \\ 
 \hline 
 dense\_2 &25&75&150&200&250&300&350&400&450&550&650&900&1150&1400&1650 \\ 
 \hline
 Parameters ($\times 10^{6}$) &0.013&0.047&0.097&0.152&0.22&0.29&0.38&0.48&0.58&0.75&0.93&1.76&2.85&4.17&5.75\\
 \hline
 \end{tabular}
\caption{Number of trainable parameters as a function of number of neurons in fully connected dense layers for two-qubit tomography.}
\label{table:one}
\end{table*}

\begin{figure}[t!]
\centering
\includegraphics[width=\linewidth]{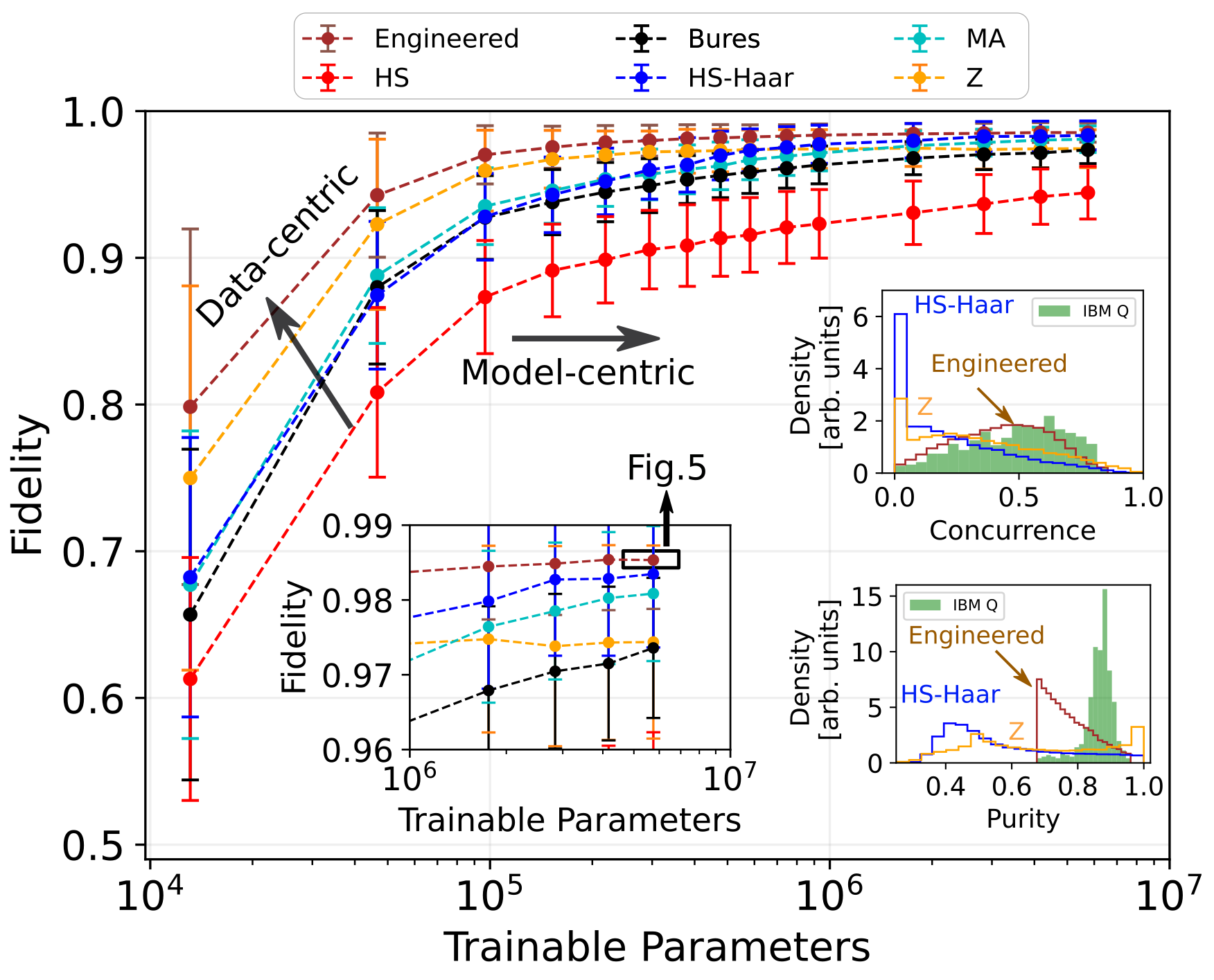}
\caption{\textbf{Reconstruction fidelity versus number of trainable parameters for various training set distributions.} The use of various distributions of random quantum states to train the network constitutes a data-centric approach and is shown by an arrow pointing upwards, whereas varying number of trainable parameters is an example of a model-centric approach as indicated by the arrow pointing to the right. The domain from $10^6$ to $10^7$ parameters  
is magnified in the left inset. The rectangular box in the inset indicates the network architecture used for the results described in Fig.~\ref{fig:fids_shots}. Additionally, the concurrence and purity of random quantum states from the Hilbert--Schmidt--Haar (HS--Haar), \.{Z}yczkowski (Z), engineered, and IBM Q distributions are, respectively, shown by blue, orange, brown, and green histograms in the right insets. 
}
\label{fig:fid_vs_params}
\end{figure}

The results of the data-centric and model-centric approaches are pictured simultaneously in Fig.~\ref{fig:fid_vs_params}.
Each curve corresponds to a different data-centric method, meaning the neural network was trained with states drawn from a different distribution.
For each curve, increasing along the $x$-axis corresponds to a ``model-centric'' performance improvement, where the training set is fixed but the number of trainable parameters in the network is increased.  
Each point making up each curve corresponds to the average reconstruction fidelity for our ML-QST system when reconstructing 500 NISQ-sampled states, using a network trained on 30,000 randomly sampled states from the corresponding distribution.  Note that all measurements  used to train the networks described in Fig.~\ref{fig:fid_vs_params} are simulated in the ideal scenario (i.e., the limit of infinite shots, which in Sec.~\ref{shotNoise} we will see is a reasonable choice given the test set is sampled at 1024 shots). 

From Fig.~\ref{fig:fid_vs_params} we see that the average performance of our system for any training set is improved, at least at first, using model-centric techniques.  However, the performance improvements from this model-centric approach appear to impact each iteration of our network in approximately the same way (with a few minor crossovers occurring).
Similarly, for a given network size ($x$-axis position) we find the average reconstruction fidelity can be improved with data-centric methods.
In other words, we find that data-centric and model-centric approaches are complementary paths to performance improvement.

Ultimately, Fig.~\ref{fig:fid_vs_params} indicates that the engineered training set attains the highest average reconstruction fidelity.  
Importantly, only the minimum and maximum values of the NISQ purity and concurrence were used in producing the training sets---not any detailed features of the distribution's shape.
The HS--Haar distribution, which is simply a convex sum of HS- and Haar-distributed states, performs close to that of the engineered distribution in the limit of the maximum model-centric improvements.
This is especially surprising when considering the inset of Fig.~\ref{fig:fid_vs_params} which shows how dramatically the engineered and HS--Haar distributions differ in both purity and concurrence.  
The impact of these differences evidently shrinks as the number of trainable parameters grows, removing the initially wide separation (at small $x$-values) in the performance of the neural networks trained on the engineered distribution and HS--Haar.

After the engineered and HS-Haar training sets the next highest performing sets are those which can be biased based on mean purity, the MA and Z distributions. 
As discussed above our selection of MA parameters were heuristically chosen based only on $P_\mathrm{min}$.
To better understand the optimality of this selection we also include the reconstruction fidelities with various MA concentration parameters in Appendix~\ref{Only_MA_distributed_results}. It is unsurprising that all of these distributions ultimately outperform the Bures and HS training sets, both because they have no parameters with which to incorporate prior information and because they skew significantly more mixed than the average state generated by $ibmq\_jakarta$.

\begin{figure}[tb!]
\centering
\includegraphics[width=.8\linewidth]{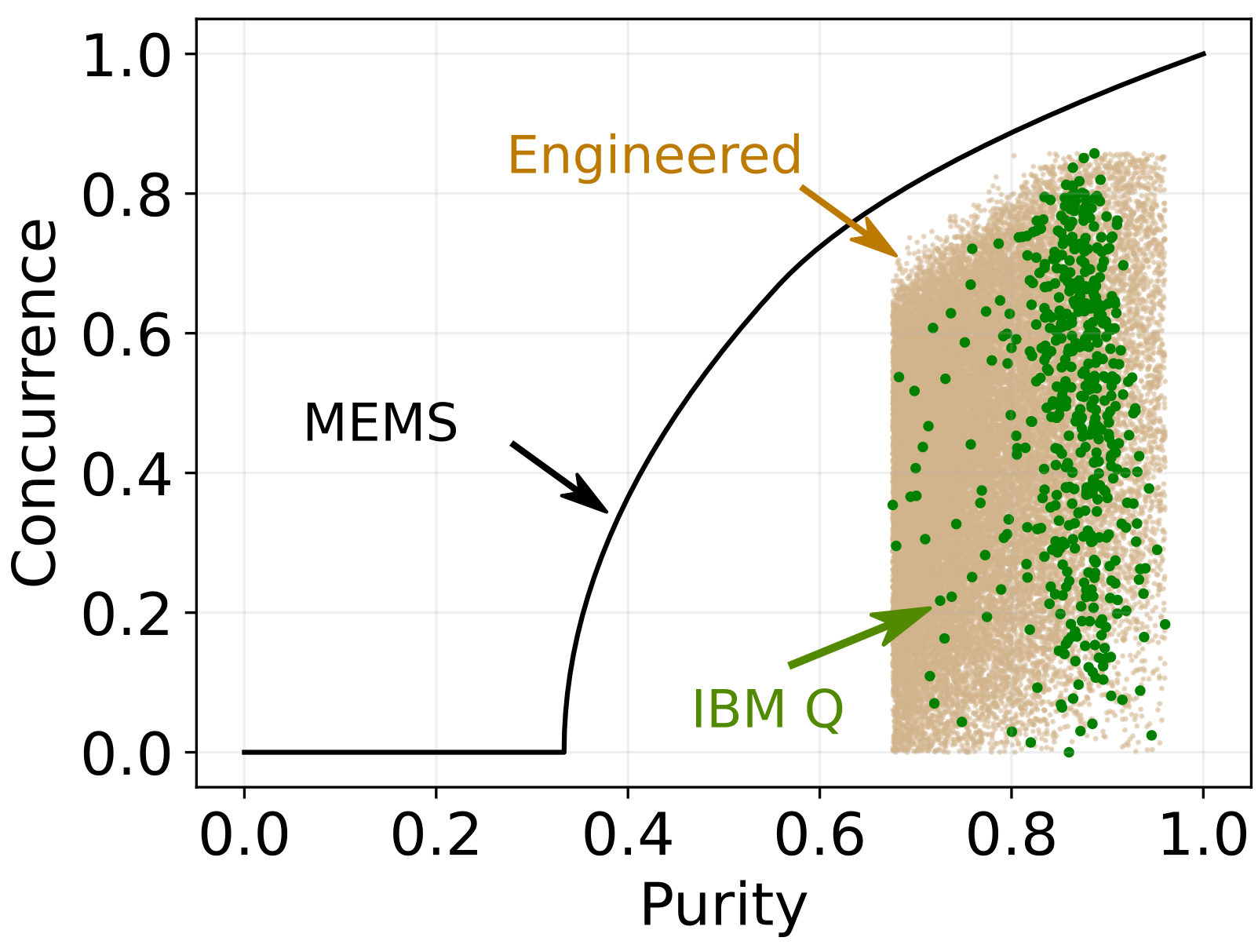}
\caption{\textbf{Engineered states on concurrence-purity plane.} The engineered and IBM Q sets are shown by brown and green dots, respectively.}
\label{fig:ibmq_eng}
\end{figure}

To better understand exactly how the engineered distribution---which again only takes into account minimum and maximum information---compares to the distribution actually generated by the NISQ device, in Fig.~\ref{fig:ibmq_eng} we plot  the concurrence of the states sampled as a function of purity from  $ibmq\_jakarta$ (labeled as IBM Q in green) and those from the engineered distribution (in brown).
We see that the engineered set covers the NISQ-sampled set convincingly, albeit weighted more heavily toward mixed states.
While this might initially appear to be an inefficiency, we will find in Sec.~\ref{heterogeneity} that this bias in the training set toward lower purity than the target distribution actually helps explain the high performance of the engineered set.

\begin{figure}[b]
\centering
\includegraphics[width=\linewidth]{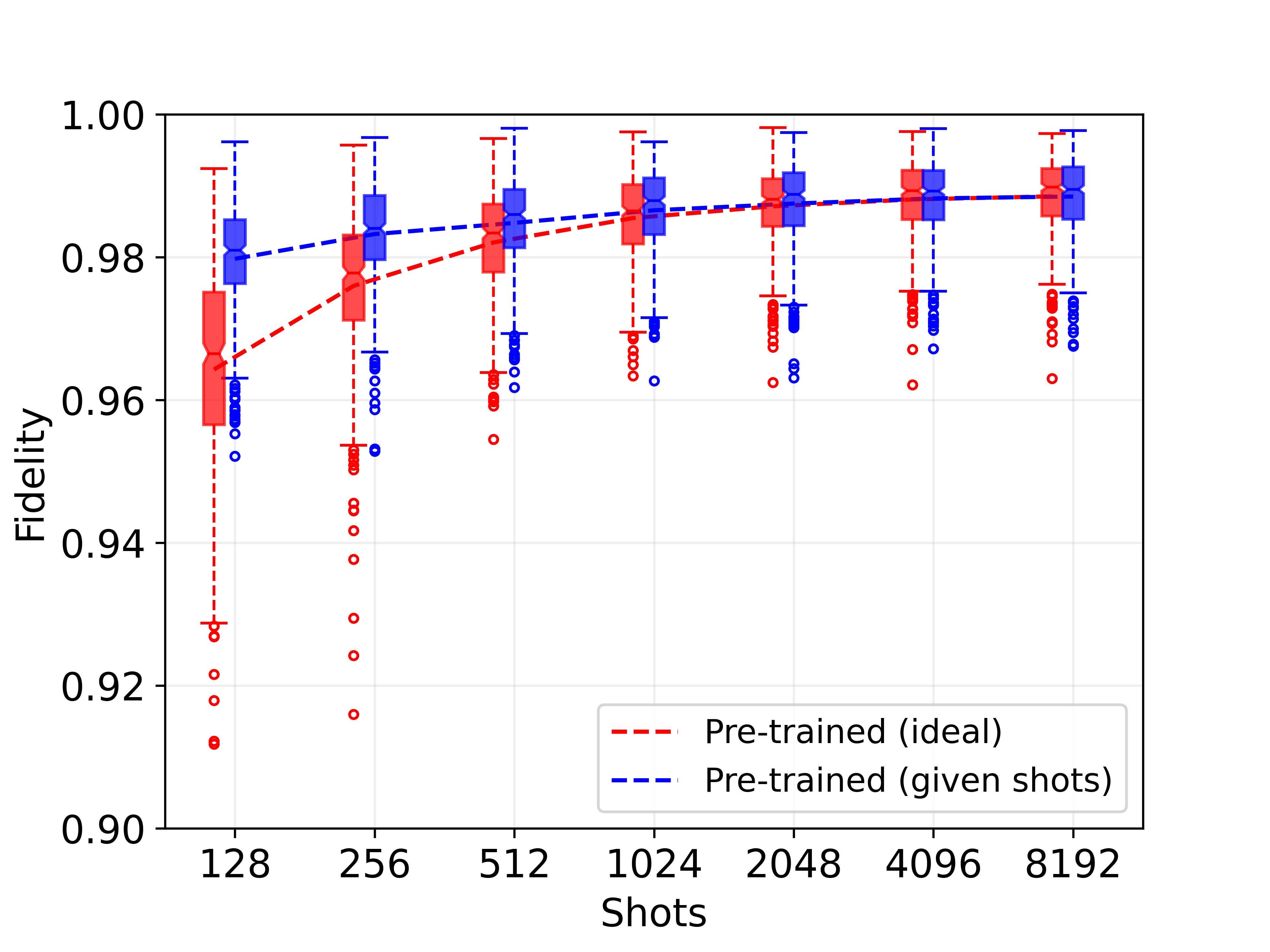}
\caption{\textbf{Data-centric approach in the low-shot regime.} Reconstructing the NISQ-sampled distribution of Sec. \ref{V_A} with simulated measurements performed  with shots ranging from 128 to 8192. The red line is the reconstruction fidelity when performed using a network trained on ideal measurements which themselves have no statistical error.  The blue line is the reconstruction fidelity when a separate network has been trained for each shot level such that the training set was simulated at the same shot level as the test set. The whiskers represent the inter-quartile range, while small circular dots are the outliers.}
\label{fig:fids_shots}
\end{figure}

\begin{figure*}[ht]
\centering
\includegraphics[width=\linewidth]{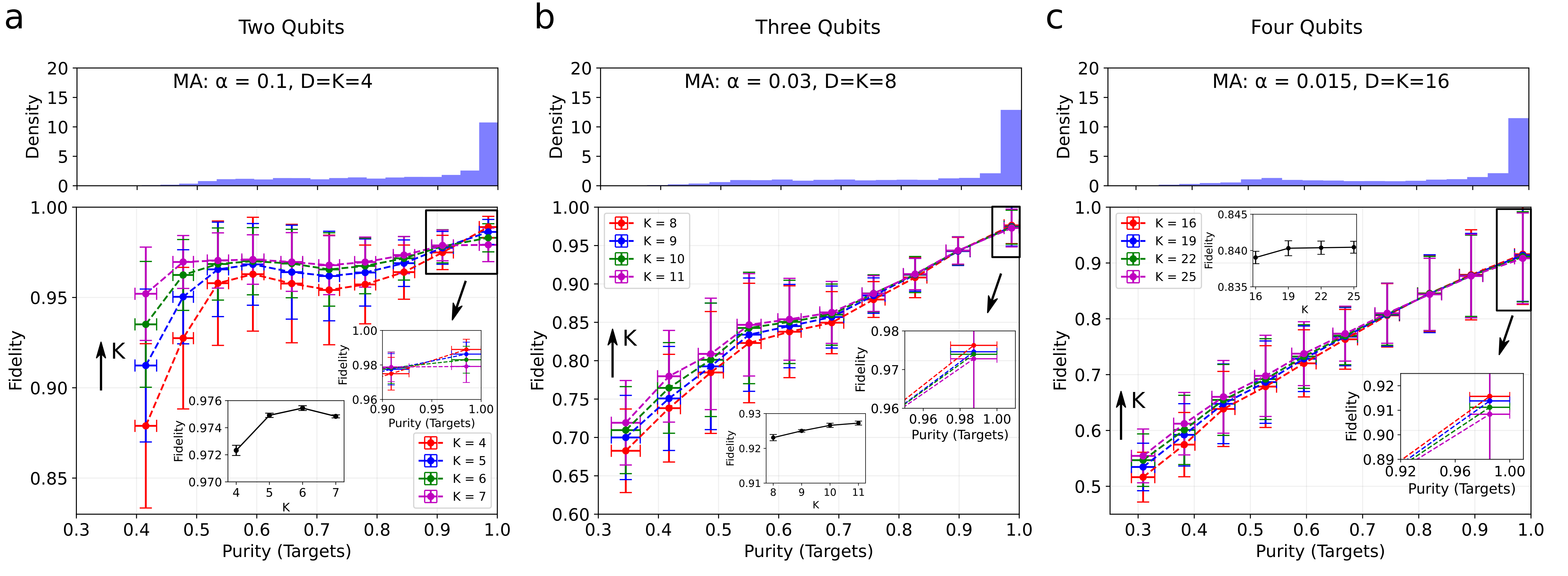}
\caption{\textbf{Heterogeneous state complexity.} Reconstruction fidelities versus test state purity for (a) two-qubit, (b) three-qubit, and (c) four-qubit quantum systems. The histograms for the purity distributions of the target states are shown at the top. In the lower plots, the red lines indicate reconstruction when the network has been trained using the same distribution (sampled separately) as the test set.  Each other curve (blue, green, purple) represents the reconstruction fidelity when the network is instead trained on a more mixed version of the test distribution. The vertical and horizontal error bars represent one standard deviation from the mean. Moreover, the reconstruction fidelities averaged over all test states are shown in the insets, for each $K$.}
\label{fig:fids_purity}
\end{figure*}

\section{Applications of data-centric engineering}

In this final section, we present two additional data-centric techniques for improving the reconstruction fidelity of our system. The first subsection considers situations where statistical noise is present in measurement results and demonstrates that synthetic statistical noise in training sets can significantly improve average reconstruction fidelity. The second subsection describes a surprising result applicable to scenarios where the states composing a test set vary widely in purity. In this case, even given complete access to the distribution of the test set and using that exact distribution to generate the training set, the optimal average reconstruction fidelity is not found by constructing a training set from the same distribution but rather from one slightly more mixed. The two methods in this section can be used independently or in concert with each other and the other heuristics described throughout this paper.  

\subsection{Low-shot state reconstruction}
\label{shotNoise}

Experimental data used for state reconstruction will necessarily include statistical noise since measurements can only be repeated a finite number of times. 
Many practical considerations may further restrict the plausibility of repeating an experiment, such as low count rates, experimental complexity, or the dimension of the underlying quantum system, which results in inefficient scaling of required measurements.
The presence of statistical noise in measurement results causes estimated expectation values of measurement operators to differ from the ideal, lowering the reconstruction fidelity.  
In the context of ML-based QST systems, some previous work has demonstrated that incorporating statistical noise comparable to that present in a test set into the training set can improve average reconstruction fidelity of pure states \cite{lohani2021experimental}.
Here, we extend this fundamentally data-centric technique by applying it both to mixed states generally and to the engineered distribution of Sec.~\ref{V_C} specifically.
In other words, we show that multiple data-centric techniques can be used in a complementary fashion. 

For our demonstration we use the same states obtained  in Sec.~\ref{V_A} as our test set, but with their measurements simulated at shots ranging from 128 to 8192.
Here the term ``shot'' represents the number of times each Pauli measurement circuit runs. 
In Fig.~\ref{fig:fids_shots}, we show the reconstruction fidelity of the NISQ-sampled (from IBM Quantum) test set as a function of the number of shots used to generate the measurement results in the test set for two different training sets.
We use the same network architecture, with $dense\_1\,=\,3050$ and $dense\_2\,=\,1650$, with each training set.
The red line indicates the reconstruction fidelity of the test set using a network trained on ideal measurement data, meaning we generate measurement probabilities directly from expectation values.
Alternatively, the blue line is the reconstruction fidelity when the network has been trained on measurement results that have been simulated at the same shot number as the NISQ-sampled test set 
($x$-axis).  

For each data point on Fig.~\ref{fig:fids_shots}, box plots represent the range of reconstruction fidelities of the test states for the respective case. The notch indicates the median; the whisker and each box encloses $[Q_1,\, Q_3]$; and the whiskers range from $Q_1 - 1.5(Q_3-Q_1)$ to $Q_3 + 1.5(Q_3-Q_1)$, where $Q_1$ and $Q_3$ are the first and third quartiles. As can be seen by the divergence of the red and blue lines at low shot numbers, when significant statistical noise is present in a test set it is advantageous to include equivalent statistical noise in the training set.
We note that the convergence of the blue and red lines is expected as 8192 shots is large enough to significantly reduce statistical noise.


\subsection{Accounting for heterogeneity in state complexity}
\label{heterogeneity}

An intuitive assumption when generating a training set is that one should aim to match the distribution of the test set as much as possible.
Surprisingly, we will find here that it is not always optimal to exactly match the test distribution when the states cover a wide range of purity.
In particular, we find that a higher reconstruction fidelity is obtained for a given test distribution when we train our network on a slightly more mixed distribution.
We will demonstrate this effect for two, three, and four qubit systems.

To control the relative mean purity of the test and training distributions, we use the MA distribution as described in Sec.~\ref{distributions}. 
We begin by fixing the test distribution concentration parameter and $K$ value.
We then train a network on the same distribution as well as several others with the same concentration parameters but progressively larger $K$ values.
Recall that the MA distribution is constructed by summing $K$ Haar-random pure states, and hence increasing $K$ causes the overall distribution to become more mixed \cite{lohani2021improving}.

For our test sets, we draw 5000 random quantum states from the MA distribution with the parameters $(\alpha,K)=(0.1,4)$ for two qubits, $(\alpha,K)=(0.03,8)$ for three qubits, and $(\alpha,K)=(0.015,16)$ for four qubits, simulating ideal Pauli measurements on each.
Note that for informationally complete tomography, the number of measurements and number of neurons in $dense\_3$, respectively, scales as $6^d$ and $2^{2d}$, where $d$ is the number of qubits. The purity distributions of the samples for each case is, respectively, shown at the top in Fig.~\ref{fig:fids_purity}(a--c).  At each qubit number, we generate four training sets each with 30,000 random quantum states sampled from MA distributions with the same $\alpha$ as the corresponding test set, but with varying $K$: $K\in\{4,5,6,7\}$ for two qubits, $K\in\{8,9,10,11\}$ for three qubits, and $K\in\{16,19,22,25\}$ for four qubits.

We train a separate network at $dense\_1\,=\,3050$, and $dense\_2\,=\,1650$ for each $K$ and reconstruct the test states. To collect statistics, we run each network 10 times and take the average of all 10 predictions for each test state as the reconstruction fidelity for the given state. The reconstruction fidelities, grouped by the ground truth purity of the test states, are plotted in Fig.~\ref{fig:fids_purity}(a) for two qubits, (b) for three qubits, and (c) for four qubits. The purity range from 0.3 to 1.0 is divided into 10 bins, and the statistics are evaluated separately in each bin. The vertical and horizontal error bars represent one standard deviation from the binned mean of the reconstruction fidelity and ground truth purity, respectively. 

In general, we find that increasing $K>D$ in the training set, which decreases the purity, significantly enhances reconstruction fidelities for mixed states, while slightly reducing performance for pure states (as shown in the insets of Fig.~\ref{fig:fids_purity}).
Therefore, caution should be taken when choosing the value of $K$ used in the generation of the training set. Nevertheless, on the whole, the improvement for mixed states tends to outweigh any reduction in performance for  pure states.  We conjecture that this effect can be explained by the difference in the number of terms required to fully describe a state of different purity.  For example, a pure state has fewer free variables than a mixed state of the same dimension, making it more difficult for the network to learn how to reconstruct a mixed state than a pure state. Hence, biasing a training set to be slightly more mixed than the target distribution improves the performance of the network on average.  
    
\section{Discussion}
\label{discussion}
Data-centric techniques represent a broad set of valuable and often underutilized strategies for improving the performance of classical ML-based systems used throughout QIS.  Unlike model-centric approaches, data-centric methods have the distinct advantage of requiring no alteration to the underlying ML model.  Generally speaking, data-centric techniques focus on identifying inadequacies in the construction of data sets, such as false correlations, insufficient variety of examples, and improper scoping. Remedying these deficiencies can significantly improve the performance of ML-based systems, but identifying these errors can require significant domain-specific knowledge.  This paper has developed various data-centric heuristics for training set generation that consider prior or domain-specific knowledge to improve system performance, demonstrating the effectiveness of these heuristics with an ML-based quantum state reconstruction system. 

Many data-centric heuristics are highly specialized to a particular situation under investigation and broadly include any technique for incorporating prior knowledge, such as the expected average state a system will generate, into the structure of the generated data set.  Previous work has considered how to create data sets for ML-based quantum state reconstruction that take into account statistical counting noise, systematic experimental errors, and the expected distribution of states generated by a system \cite{lohani2021experimental, lohani2021improving}.  Here, we have added to this list a method for engineering training sets to match distributions of expected experimental scenarios.  We compare the effectiveness of our distribution-engineering approach to other standard methods for generating data sets, including those capable of incorporating some amount of prior knowledge such as mean purity.  

We describe how spurious correlations can reduce system performance, how it can be challenging to identify these correlations in quantum states given their complexity, and how the inclusion of only a few counterexamples can remedy problems related to these correlations.  We show that even for systems as small as two qubits it can be tempting to believe a data set is broadly illustrative of the overall set of possible states.  In particular, we generate a data set that includes nearly the full range of possible purity and concurrence values and yet contains a false correlation between the two.  We show that, in this example, such a correlation causes our state reconstruction system to misclassify pure separable states as entangled, having only ever seen pure states that are entangled.  We then demonstrate that surprisingly few counterexamples need to be added to the training set to remedy this issue.  Hence, it is prudent to include several states of every possible classification in any given data set.

More generally, we have also described data-centric heuristics that leverage only broad features of QIS rather than specific prior knowledge about an experiment or scenario.  In particular, we find that, given the heterogeneity between the number of free-variables in pure and mixed states, it is not always optimal to endeavor to generate training sets that exactly match the distribution of an experimental scenario in the first place. Instead, when an ML system is to be applied to states covering a wide range of purities, training sets should be biased to be more mixed on average than the expected experimental distribution. 

The data-centric heuristics described in this paper focus on situations where training data are synthetically generated, as is often the case in applications of classical ML to QIS-specific problems.  The motivation for simulated data sets can be due to convenience, as experimental data may be impractical to obtain, or because the problem itself is theoretical and measured data would only open the possibility of introducing experimental errors.  However, the heuristics developed still apply to experimentally obtained training sets, but in those cases can be considered more prescriptive as they suggest the structure of data sets likely to result in the highest performing ML systems.   Due to our focus on synthetic data, we have not included any data-centric methods concerned with data labeling. However, we note that significant work in the ML community has focused on the effects of data labeling and developed a set of data-centric approaches for systematically relabeling or removing mislabeled data to improve overall system performance \cite{pan2022data, ngiam2019big, tanaka2018data, huang2021power}.  An interesting problem for future studies would be to consider the application of these label-focused approaches to experimental QIS systems.

\begin{figure*}[t]
\centering
\includegraphics[width=\linewidth]{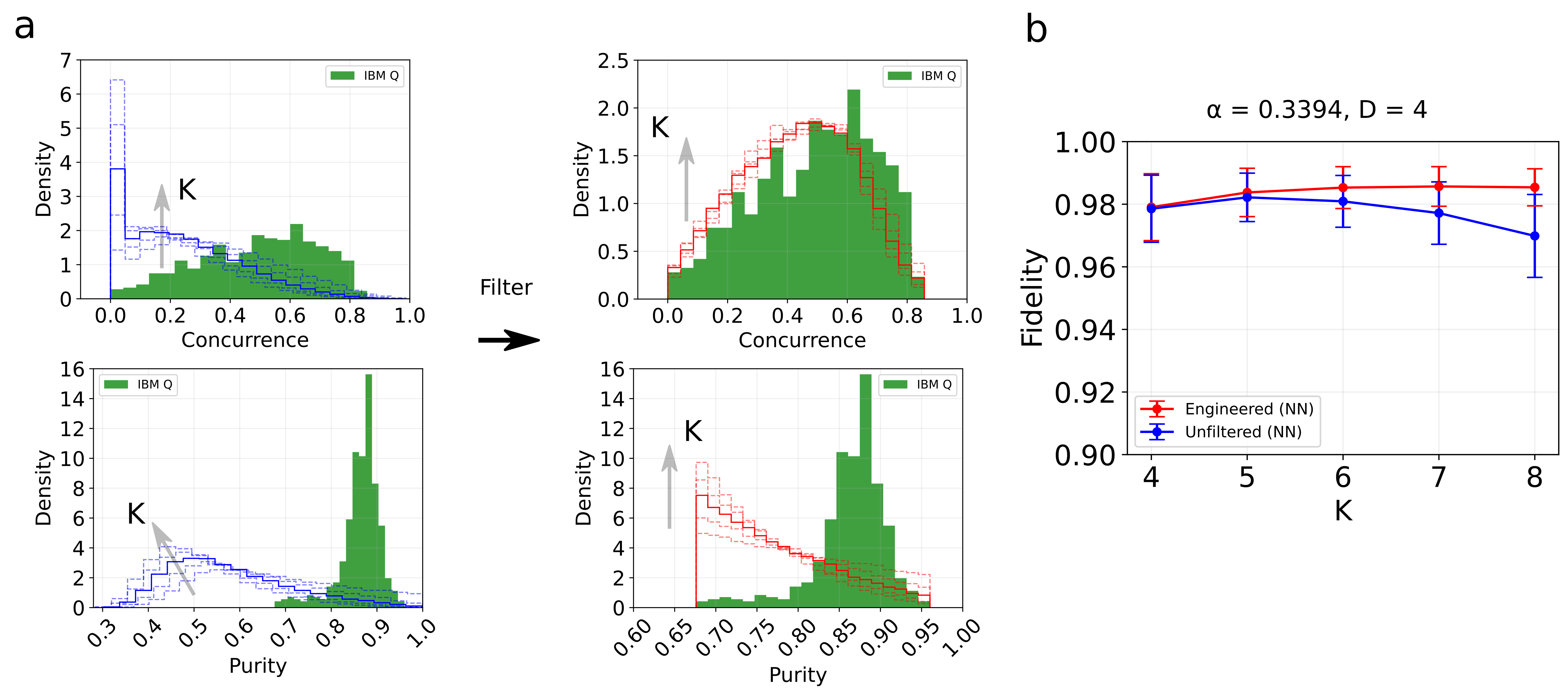}
\caption{\textbf{Engineered states.} (a) Unfiltered (left) and  engineered (right) state distributions from MA with $\alpha = 0.3394$ and $D = 4$, with respect to concurrence (top) and purity (bottom). The value of $K$ consecutively increases in each plot from 4 to 8. (b) Reconstruction fidelities versus the value of $K$.}
\label{fig: fid_vs_k_appendix}
\end{figure*}

\begin{acknowledgments}
Work by S. Lohani and T. A. Searles was supported in part by the U.S. Department of Energy, Office of Science, National Quantum Information Science Research Centers, Co-design Center for Quantum Advantage (C2QA) under contract number DE-SC0012704. A portion of this work was performed at Oak Ridge National Laboratory, operated by UT-Battelle for the U.S. Department of Energy under contract no. DE-AC05-00OR22725. J.M.L. acknowledges funding by the U.S. Department of Energy, Office of Science, Advanced Scientific Computing Research, through the Early Career Research Program.  The views and conclusions contained in this document are those of the authors and should not be interpreted as representing the official policies, either expressed or implied, of the Army Research Laboratory or the U.S. Government. The U.S. Government is authorized to reproduce and distribute reprints for Government purposes notwithstanding any copyright notation herein. Additionally, we acknowledge use of the IBM Quantum for this work. The views expressed are those of the authors and do not reflect the official policy or position of IBM Quantum. This material is based upon work supported by, or in part by, the Army Research Laboratory and the Army Research Office under contract/grant numbers W911NF-19-2-0087 and W911NF-20-2-0168.
\end{acknowledgments}


\appendix

\section{Varying $K$}\label{appendix_C}
Due to asymmetry in state complexity, the performance of a neural network may depend on the density of mixed and pure states in a training set. First, in order to generate a more mixed training set, we vary the value of $K$ consecutively from 4 to 8 using the MA distribution at $\alpha=0.3394$ and $D=4$, and the corresponding distributions of the states with respect to the concurrence and purity are shown by unfilled blue histograms in Fig.~\ref{fig: fid_vs_k_appendix}(a) (left). 
Note, this $\alpha$ value was chosen as it aligns the mean of the MA distribution with the minimum purity value in the NISQ-sampled data set, which we found in Sec.~\ref{V_C} worked well for engineered states.
The solid-line histogram represents the case of $K\,=\,6$, whereas the the filled green histogram represents the test distributions obtained from the cloud-accessed hardware (IBM Q). As shown, the increase in the value $K$ is directed upward, increasing the mixedness of the training samples. Without any filter, we find that the increase in reconstruction fidelities with $K$ quickly saturates and then gradually decreases as shown by the blue line in Fig.~\ref{fig: fid_vs_k_appendix}(b). The error-bars show one standard deviation from the mean.

In order to address the issue, we apply a filter of concurrence and purity to remove unwanted mixed states from the training set. The concurrence (top) and purity (bottom) histograms for the filtered (engineered) distributions of sampled states are shown by unfilled red histograms in Fig.~\ref{fig: fid_vs_k_appendix}(a) (right). As previously mentioned, a solid-line histogram indicates $K=6$. With a network pre-trained with these engineered states, we find the reconstruction fidelity gradually increases as shown by red line in Fig.~\ref{fig: fid_vs_k_appendix}(b), 
saturating by around $K=6$. 

\FloatBarrier
\section{Learning rate}\label{appendix_D}
\begin{figure}[!h]
\centering
\includegraphics[width=\linewidth]{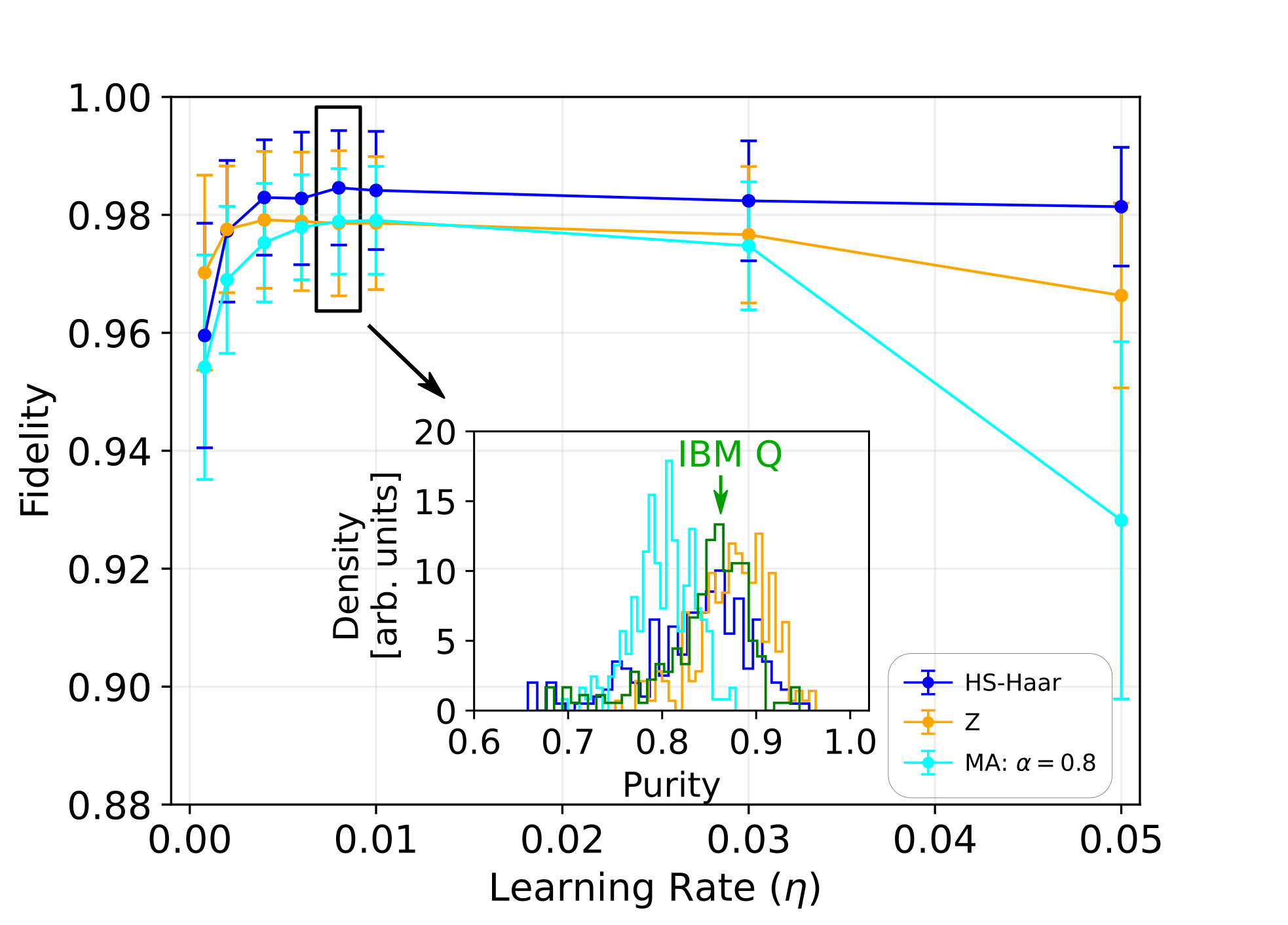}
\caption{\textbf{Optimizing learning rate.} Fidelity of reconstructed density matrices versus learning rate $\eta$. The full purity distributions of the reconstructed states for $\eta=0.008$ 
are shown in the inset. The error bars show one standard deviation from the mean fidelity.}
\label{fig:eta_vs_purity}
\end{figure}

The learning rate ($\eta$) is an important hyperparameter affecting the training of a network \cite{bengio2012practical}. We vary the learning rate for a network from 0.0008 to 0.05 and evaluate the fidelity of the reconstructed test density matrices from IBM Q. The results are shown in Fig.~\ref{fig:eta_vs_purity}. In order to optimize the rate, we use three quantum state distributions: HS--Haar (blue), Z (orange), and MA at $\alpha=0.8$ (cyan) at $D\,=\,K\,=\,4$ to train a network. We use a separate network for each training set and vary the learning rate. We find that increasing the rate parameter gradually increases the fidelity for all the training cases and peaks around $\eta = 0.008$. Additionally, we show the purity distributions of the reconstructed density matrices when $\eta=0.008$ in the inset. We find that at $\eta = 0.008$ the predicted purities have good overlap with the target IBM Q distribution.

\FloatBarrier
\section{Reconstructing states with MA-distributed training sets}\label{Only_MA_distributed_results}

In this Appendix, we consider the performance of our heuristically chosen values for $\alpha$ and $K$ in the MA distribution of Fig. \ref{fig:fid_vs_params} by reproducing this curve along with several other parameter choices.  We see that our heuristic choice performs slightly worse for models with fewer trainable parameters but ultimately outperforms the different curves as the trainable parameters increase.

\begin{figure}[!h]
\centering
\includegraphics[width=\linewidth]{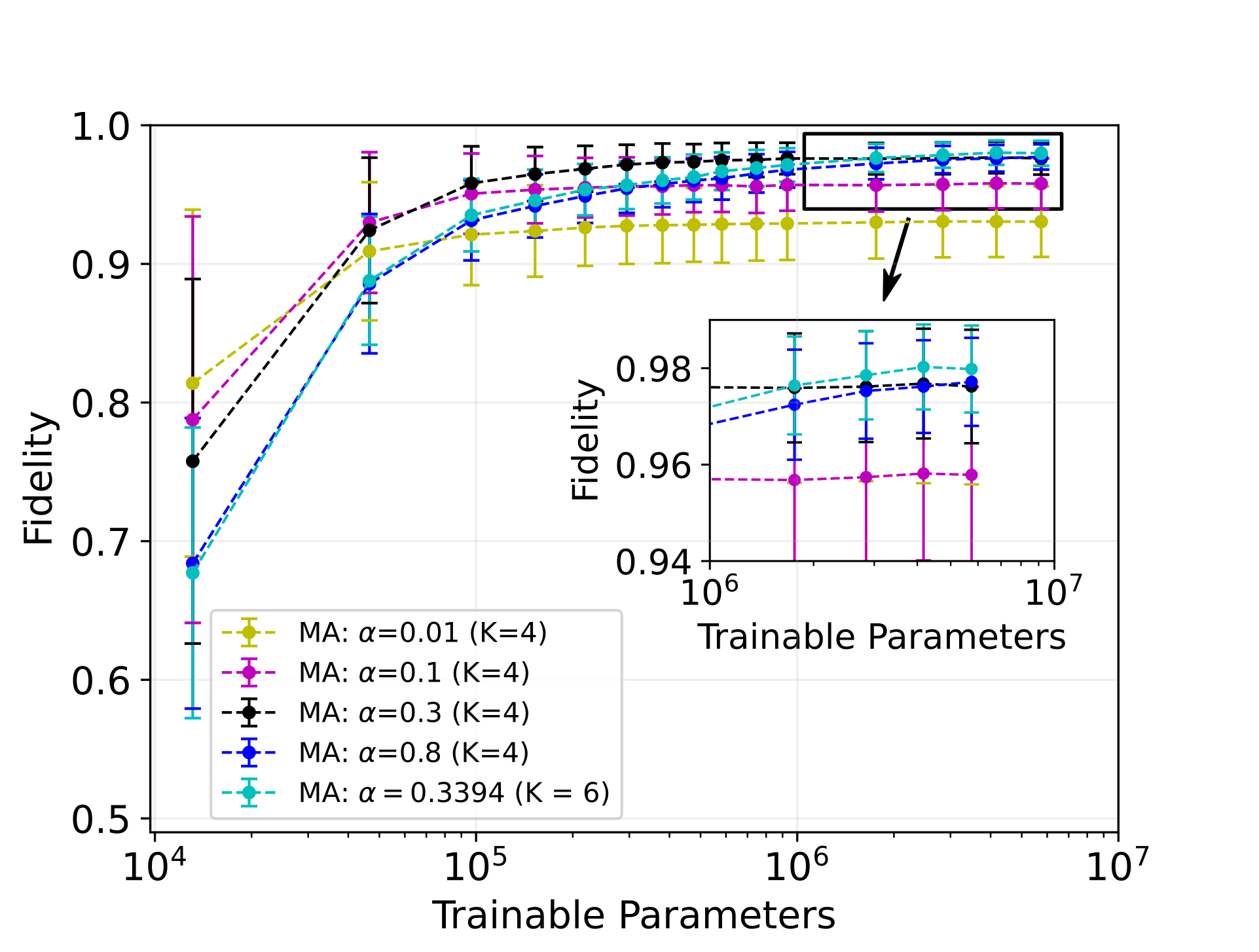}
\caption{\textbf{Reconstruction fidelity versus trainable parameters for various MA-distributed training sets.} The pairs of concentration parameter and K-value are chosen as $(\alpha, K) \in \{(0.01, 4), (0.1, 4), (0.3, 4), (0.8, 4), (0.3394, 6)\} $ for training sets. The results shown by a dotted cyan-line is derived from Fig. \ref{fig:fid_vs_params}. The error bars show one standard deviation from the mean fidelity.}
\label{fig:fid_vs_train_MA}
\end{figure}

\FloatBarrier
\section{Tuning the mean of MA-distributed training states}\label{tuning_MA}

In Sec.~\ref{V_B} we suggested setting the mean of the initial distribution, which is then sent through the bandpass filter, to $P_\mathrm{min}$.  Here we consider the performance of this parameter selection.  To this end, we generate engineered training sets where 
the initial concentration parameter $\alpha$ is chosen such that an MA distribution with $K=4$ will have mean purity given by the $x$-axis in Fig.~\ref{fig:fid_vs_train_MA}. (The mean of the engineered and $K=6$ training sets will thus differ from this value.)
We then use these engineered data sets to reconstruct the same NISQ sampled data as in Fig. \ref{fig:fid_vs_params}.  For comparison, we also perform the same exercise for MA distributions that have not gone through bandpass filters.  We see that the engineered distribution outperforms the base MA distribution in all cases.  Further, our heuristic choice of setting the mean of the MA distribution to $P_\mathrm{min}$ is near-optimal over the presented range.  

\begin{figure}[!h]
\centering
\includegraphics[width=\linewidth]{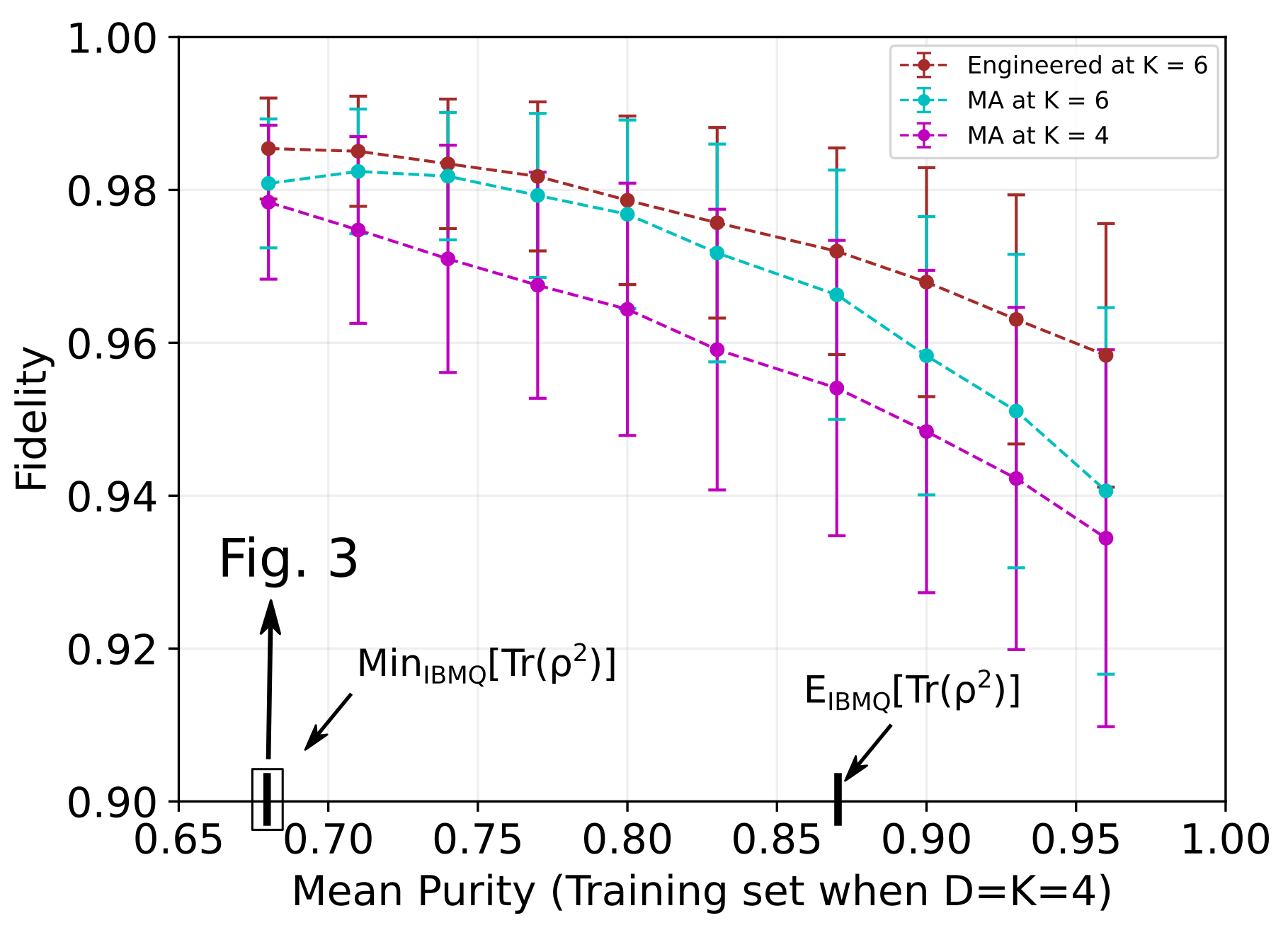}
\caption{\textbf{Reconstruction fidelity of NISQ-sampled test set versus the mean purity of various MA-distributed training states when $K=4$.} 
The two vertical tick marks along the $x$ axis emphasized with arrows correspond to the scenarios where the mean purity of the training set matches the minimum and mean purity of the NISQ sampled states when $D\,=\,K\,=\,4$. 
The mean purity of the states (labeled with subscripts IBMQ) is shown by a vertical line on the $x$-axis as indicated by an arrow. The error bars represent one standard deviation from the mean.}
\label{fig:fid_vs_train_MA}
\end{figure}

\FloatBarrier
\section{Engineered distributions}\label{appendix_E}

Here we present the complete algorithm written in pseudocode for generating engineered distributions of quantum states as explained in Section \ref{V_B}.

\newcommand\mycommfont[1]{\footnotesize\ttfamily\textcolor{blue}{#1}}
\SetCommentSty{mycommfont}
\SetKwInput{KwInput}{Input}                
\SetKwInput{KwOutput}{Output}              
\DontPrintSemicolon
\begin{algorithm}[h]  
    \KwInput{ $P_\mathrm{min}$,$P_\mathrm{max}$, $C_\mathrm{min}$, $C_\mathrm{max}$ }
    \KwOutput{Engineered States ($\rho_{eng}$)}
        {$\alpha\, \gets \,\frac{D\Big[1\,-\,P_\mathrm{min}\Big]}{D\Big[D\,P_\mathrm{min}\,-\,1\Big]\,-D\,+\,1}$ \\
        $\balpha \gets (\alpha,\, \alpha,\, \alpha,\, \alpha, ..., \alpha) \; (\textrm{K terms})$};\, $K \ge D$
    
    \vspace{10pt}
    \tcp{Draw N samples}
    $[\bx]_{N, K} \gets \text{Dir}(x\vert\alpha)=\frac{\Gamma\left(\sum_{i=1}^{K}\alpha_{i}\right)}{\prod_{i=1}^{K}\Gamma(\alpha_{i})}\prod_{i=1}^{K}x_{i}^{\alpha_{i}-1}$\\
    
    \vspace{10pt}
    \tcp{Reshape into a 4D-array}
    $[x]_{N, K, 1, 1} \gets \bx_{N, K}$\\
    Generate $N \times K$ Haar-random pure-states, $[\rho_\mathrm{Haar}]_{N\times K, 4, 4} \gets |\psi\rangle\langle\psi|$\\
    $[\rho_\mathrm{Haar}]_{N,K,4,4} \gets [\rho_\mathrm{Haar}]_{N\times K, 4, 4}$\\
    
    \vspace{10pt}
    \tcp{Perform the element-wise multiplication, and then, take a sum along the first dimension}
    $[\rho_{MA}]_{N,4,4} = \sum_{k=1}^K [x]_{N,\, k,\, 1,\, 1}\cdot [\rho_\mathrm{Haar}]_{N,\, k,\, 4,\, 4}$ \\

    \vspace{10pt}
    \tcp{Collect N Pauli Y-Operators and reshape it to a 3D-array}
    $[Y]_{N, 2, 2} \gets Y_N$\\
    $[\tilde{\rho}_{MA}]_{N,4,4} \gets [\rho_{MA}^{*}]_{N,4,4}$\\
    $[R]_{N, 4, 4} \gets \sqrt{\sqrt{\rho_{MA}}(Y\otimes Y)\tilde{\rho}_{MA}(Y\otimes Y)\sqrt{\rho_{MA}}}$\\
    $[\lambda]_{N,\,4} \gets eigen\_values([R]_{N, 4, 4})$\\
    
    \vspace{10pt}
    \tcp{Sort eigen values in increasing order, $\lambda_1 > \lambda_2 > \lambda_3 > \lambda_4$}
    $\Big[(\lambda_4,\, \lambda_3,\, \lambda_2,\, \lambda_1)\Big]_{N, 4} \gets sort\Big([\lambda]_{N,\,4}\Big)$\\
    
    \vspace{10pt}
    \tcp{Evaluate the concurrence $(C)$ and purity $(P)$ of $\rho_{MA}$ as}
    $[C(\rho_{MA})]_N \gets \textrm{max}(0,\, \lambda_1\, -\, \lambda_2\, -\, \lambda_3\, -\, \lambda_4)$\\
    $[P(\rho_{MA})]_N \gets \Tr([\rho^2_{MA}]_{N, 4, 4})$\\
    
    \vspace{10pt}
    \tcp{Perform the element-wise logical AND operation}
    $\rho_\mathrm{eng} \gets \rho_{MA}\Big[\big(C_\mathrm{min}\leq [C(\rho_{MA})]_N \leq C_\mathrm{max}\big)\, \&  \big(P_\mathrm{min}\leq [P(\rho_{MA})]_N \leq P_\mathrm{max}\big)\Big]$\\
    \textbf{return} $\rho_\mathrm{eng}$\\
    \tcc{End}
\caption{Engineered distributions}
\end{algorithm}

    
    
    
            

\newpage

\bibliographystyle{apsrev4-1}
\bibliography{refs-ML}

\end{document}